\documentclass{scrartcl}
\usepackage{amsmath,graphicx,amssymb}
\usepackage[german,english]{babel}
\usepackage{psfrag}
\newcommand{\vp}{\varphi}
\newcommand{\pa}{\partial}
\newcommand{\rd}{r_d}
\newcommand{\al}{\alpha}
\newcommand{\Om}{\Omega}
\newcommand{\eps}{\epsilon}
\newcommand{\M}{\mathcal{M}}
\newcommand{\xd}{x_d}
\renewcommand{\vec}[1]{\boldsymbol{$#1$}}

\begin{document}

\title{Motion and gravitational radiation of a binary system consisting of
an oscillating and rotating coplanar dusty disk and a point-like object}
\author{D\"orte Hansen\footnote{e-mail:nch@tpi.uni-jena.de}\\
Theoretisch-Physikalisches Institut\\
Friedrich-Schiller-Universit\"at Jena \\
Max-Wien-Platz 1 \\
07743 Jena}
\date{}
\maketitle

\selectlanguage{english}

\begin{abstract}
A binary system composed of an oscillating and rotating coplanar
dusty disk and a point mass is considered. The conservative dynamics
is treated on the Newtonian level. The effects of gravitational
radiation reaction and wave emission are studied to leading quadrupole
order. The related waveforms are given. The dynamical evolution of
the system is determined semi-analytically exploiting the
Hamiltonian equations of motion which comprise the effects both
of the Newtonian tidal interaction and the radiation reaction on
the motion of the binary system in elliptic orbits. Tidal resonance
effects between orbital and oscillatory motions are considered in the
presence of radiation damping.
\end{abstract}

\section{Introduction}

Compact binaries  are among the most promising sources for the
gravitational waves that should be detected by gravitational wave
observatories such as GEO600, LIGO, TAMA, and VIRGO
on Earth, or LISA in space. On the other hand, it
is well known that oscillating compact objects also emit gravitational waves that
should be detectable by gravitational wave observatories. Though the
ring-down phases of merged binaries or collapsed objects are 
the strongest sources for gravitational waves from oscillating
objects, oscillations  of the
components of binary systems generated by tidal interaction
during their inspiralling phase should induce measurable characteristic modifications in the gravitational waveforms
compared with those from binaries with point-like components.
Particularly in the leading resonance cases where the orbital frequency
is a factor of two or three smaller than the oscillation frequencies,
the internal motion of the binaries is expected to be
strongly excited.\\

The treatment of the motion of compact binaries within the full
Einstein theory of gravity is a demanding challenge. Much effort has been invested in this
challenge on the numerical simulation side because an exact analytic solution
cannot be expected to be achieveable. However, a fully
numerically simulation is still far ahead. \\

The aim of the present paper is the development of a quite simple approximate
model to the full Einstein theory which shows typical features of inspiralling
compact binaries and which is fully analytic apart from the numerical
integration of a system of ordinary differential equations.
This model of a binary system is given by a point-like object
and an oscillating, rotating dusty disk, where the orbital plane and
the plane of the disk coincide. As there is no pressure involved, an
extreme simplification occurs for the field equations as well as for
the equations of motion. The dynamics is treated at the level of
Newtonian gravity augmented by the gravitational radiation reaction
dynamics, which occurs in the full Einstein theory to leading
approximation at the order $1/c^5$ ($c$ denotes
the speed of light) which is, counted in powers of $1/c^2$, 2.5 orders
beyond the Newtonian order of $1/c^0$. Within this model we further
assume that the most general motion of the dusty disk, apart from
its orbital motion, is rotation and oscillation, i.e. we do not
allow the disk to  leave this configuration. This means that
the tidal force is considered as a perturbation only. The same is true
for the  radiation reaction force which is of 2.5pN (post-Newtonian) order.

The Newtonian oscillating rotating disk of dust is interesting for it's own
reasons. While stellar oscillations are usually treated as small
perturbations, the oscillations of the disk can be arbitrarily large,
they are bulk oscillations. Moreover, the equation of motion for the
disk, which can be derived from the equations of hydrodynamics,
exhibits a surprising similarity to the well-known Keplerian
problem. Exploiting this analogy, it is possible to write
down a parametrized solution for the disk's oscillations
\cite{Hunter:1965}.\\

Having fully specified our model, let us shortly summarize
the present knowledge of the analytical motion of binary systems
in the Einstein theory of gravity now. Regarding point-like binaries without
proper spins the conservative motion is known fully analytically
up to the 3pN order, where the  2.5pN order is neglected
\cite{Memmesheimer:2004cv};
the motion of a single oscillating and rotating dusty disk in it's
rest frame is known fully analytically to 1pN order \cite{Schafer:1994a}. If the
spins are included into the binary motion, in particular, if the spin
orientations are parallel or anti-parallel to the orbital angular
momentum as in our model, the analytical solution is known
up to 2pN order \cite{Wex:1995}. Taking into account Newtonian
tidal interaction in close binary systems, several semi-analytical
investigations to the evolution of binaries have been undertaken
\cite{Alexander:1987}, \cite{Ostriker:1986}, \cite{Press:1977}.
In \cite{Ho:1998hq} and \cite{Schafer:1994a} the
semi-analytical solutions of close Newtonian binaries systems  have
been obtained, including 2.5pN gravitational damping effects.\\

The paper is organized as follows: In section II we derive the
Hamiltonian formalism for the Newtonian disk, starting from the
equations of hydrodynamics. We will not consider quadrupole radiation
damping as a secular effect, but include radiation reaction terms in
the Hamiltonian equations of motion. To that end we derive the leading
order radiation reaction Hamiltonian. Then we calculate the leading
order gravitional waveform for the isolated Newtonian disk. \\
In section III we give a short review of orbital Hamiltonian formalism
and calculate the contributions to the leading order gravitational
waveform of the orbital motion. In section IV we focus
on the binary system. The tidal interaction Hamiltonian is calculated
explicitly and given in form of a series expansion. In a next step
we derive the radiation reaction Hamiltonian for the coupled binary
system. The leading order Hamiltonian equations of motion, governing
the dynamics of the system, are solved numerically for different initial
conditions. Since the contribution of the interaction of the 
disk quadrupole with the orbital quadrupole is much smaller than all the
other contributions it can be neglected in most applications. On the
other hand, the influence of the tidal interaction potential on the
dynamics of the binary is not negligible because the exchange of orbital energy
with oscillation energy can be quite strongly.

Taking into account
leading order gravitational damping, orbital energy is dragged from
the orbit via tidal interaction. In other words, tidal interaction
speeds up the inspiral process of our binary system.

The energy
transfer is strongest in tidal resonance,
where the oscillation frequency is $n$ times the orbital frequency
($n$ ist a small integer number). We
study the effect of tidal interaction on the orbit and on the 
leading order gravitational waveform. In section V we shall discuss
our results and give an outlook on future projects.

\section{Oscillations of a rotating disk of dust -- Hamiltonian
fomulation}

One of the simplest axisymmetric systems that exhibit the essential new
features of general relativity is a thin disk of dust. During their
evolution these disks emit gravitational radiation and may eventually form
a black hole. No wonder that general relativistic disks of dust
have been subjects of interest for many years. Bardeen and Wagoner
\cite{Bardeen:1971} where the first to study 
relativistic stationary rotating disks. They succeeded to find a solution
for a rigidly rotating disk in form of a series
expansion. Unfortunately this solution can be obtained only
numerically beyond 2pN approximation. In terms of ultraelliptic
functions, Neugebauer and Meinel \cite{Neugebauer:1995pm} found an exact solution for a rigidly rotating
disk in full general relativity  using inverse scattering methods.
However, these solutions are restricted to  rigidly rotating stationary
disks. Nonstationary, rigidly rotating disks of dust were first
studied by Hunter \cite{Hunter:1965}. These disks are particularly
interesting, as they are sources of gravitational waves. In the
present paper we will focus on  oscillating MacLaurin disks. These are
infinitely thin, rotating disks of dust which are obtained as a
limiting case of MacLaurin spheroids. For these configurations
Kley and Sch\"afer \cite{Schafer:1994a} succeeded to derive a fully analytically solution up to first post-Newtonian
approximation  in a parametrized form. That solution 
exhibits a surprising similarity to the 1pN parametrized solution of
the Keplerian problem.\\
However, the study of a weakly relativistic disk as a component of a
binary system is beyond the scope of this paper. We shall restrict
ourselves on leading order results, thus considering only the
Newtonian disk. \\
Let us start by briefly reviewing the known analytic solution for the
oscillating disk, obtained from the equation of hydrodynamics for
a pressureless system.  Since we are interested in the leading order
gravitational waveform emitted by the oscillating rotating disk we
shall also discuss the related 2.5pN radiation reaction. Note
that we do not treat quadrupole radiation damping as a secular
effect. The corresponding radiation reaction terms will be included
into the Hamiltonian equations of motion.  We shall solve the
Hamiltonian equations of motion for the isolated Newtonian disk
numerically and calculate the leading
order gravitational waveform in terms of canonical conjugate variables.

\subsection{The Newtonian disk revisited}

The Newtonian dynamics of MacLaurin disks has been treated by several
authors \cite{Hunter:1965}, \cite{Shapiro:1994xg},
\cite{Schafer:1994a}, starting with the equations of hydrodynamics in
Eulerian form. Let us briefly review this standard approach before we
consider the disk in Hamiltonian formalism. \\
For an axisymmetric, pressureless configuration confined to the $z=0$
plane the equations of hydrodynamics and mass conservation in
cylindrical coordinates $(r,\vp)$ read
\begin{align}
\label{disk1}
     \frac{\pa\Sigma}{\pa t}+v_r\frac{\pa\Sigma}{\pa
     r}+\frac{\Sigma}{r}\frac{\pa}{\pa r}(rv_r)=&0, \nonumber \\
     \frac{\pa v_r}{\pa t}+v_r\frac{\pa v_r}{\pa r}-\frac{\pa U}{\pa
     r}-\frac{v_\vp^2}{r}=&0,\\
     \frac{\pa v_\vp}{\pa t}+v_r\frac{\pa v_\vp}{\pa
     r}+\frac{v_rv_\vp}{r}=&0. \nonumber
\end{align}
Here $\Sigma(r,t)$ denotes
the surface density and $v_r$ and $v_\vp$ are radial and azimuthal
velocities, respectively. The gravitational potential $U$ is the
solution to Poisson's equation
\begin{eqnarray}
\label{disk2}
     \Delta U=-4\pi G\Sigma(r,t)\delta(z).
\end{eqnarray}
For $\Sigma, v_r, v_\vp$ we shall choose an ansatz where the surface
density is that of a MacLaurin disk,
\begin{align}
\label{disk3}
     \Sigma(r,t)=\sigma(t)\sqrt{1-\frac{r^2}{\rd(t)^2}},\qquad
     v_r(r,t)=-r f(t),\qquad
     v_\vp(r,t)=r\Om(t),
\end{align}
where $\sigma(t)$ is the surface density at the center and $\rd(t)$
denotes the disk radius at a given time. The disk is rigidly
rotating with angular velocity $\Omega(t)$.  
Using this ansatz, it is possible to solve (\ref{disk1})
analytically. To that end we note that the Poisson equation
(\ref{disk2}) can be solved explicitly inside the disk. Inserting (\ref{disk3})
we find
\begin{eqnarray}
\label{disk4}
      U(r,t)=G\sigma(t)\rd(t)\frac{\pi^2}{4}\left[2-\left(\frac{r}{\rd(t)}\right)^2\right],\qquad
      (r\le \rd, z=0).
\end{eqnarray}
Substituting equations (\ref{disk3}) and (\ref{disk4}) into
(\ref{disk1}) we can derive a system of four differential equations
for the unknown variables $r_d, \sigma, f$ and $\Om$:
\begin{align}
\label{disk5}
      \dot{\sigma}-2f\sigma=&0, \nonumber \\
      \dot{r}_d+f\rd=&0,\nonumber \\
      \dot{\Om}-2f\Om=&0,\\
      -\dot{f}+f^2+\frac{G\pi^2\sigma}{2\rd}-\Om^2=&0. \nonumber
\end{align}
This system is integrated with respect to the initial conditions
$f(0)=0$ and $\rd(0)=R_d$. Combining the first two equations, one finds
\begin{eqnarray}
\label{disk6}
    \sigma(t)\rd^2(t)=\frac{3}{2\pi}M_d,
\end{eqnarray}
where $M_d$ is the total Newtonian mass of the disk. The third 
equation in (\ref{disk5}) gives
\begin{eqnarray}
\label{disk7}
     \Omega(t)\rd^2(t)=\Omega_0R_d^2,\qquad \Omega_0=\Omega(0)
\end{eqnarray}
which is constant at the Newtonian level. The remaining equation
containing $\ddot{r}_d$ reads
\begin{eqnarray}
\label{disk8}
     \ddot{r}_d+\frac{2C}{\rd^2}-\frac{h^2}{\rd^3}=0,
\end{eqnarray}
where 
\begin{eqnarray}
\label{disk9}
      C:=G\sigma(t)\rd(t)^2\frac{\pi^2}{4} = \frac{3\pi}{8}GM_d
\end{eqnarray}
and
\begin{eqnarray}
\label{disk10}
     h^2:=2C\xi^2R_d=\Omega_0^2 R_d^4.
\end{eqnarray}
$\xi^2$ is defined by $\Omega_0$ as follows,
\begin{eqnarray}
       \Om_0^2=\frac{2C\xi^2}{R_d^3}.
\end{eqnarray}
Equation (\ref{disk8}) exhibits a surprising similarity to the radial
equation in the Keplerian problem. Exploiting this similarity, the
solution to equation (\ref{disk8}) can be given in parametrized form, 
\begin{eqnarray}
\label{disk11}
       \rd=a_d(1-\eps\cos\vp),\qquad
       \frac{2\pi}{P}t=u-\eps\sin u -\pi,
\end{eqnarray}
where the definitions 
\begin{eqnarray*}
     a_d=\frac{R_d}{1+\eps}
\end{eqnarray*}
and $\eps=1-\xi^2$ hold. For not being confused with the orbital eccentricity
$e$, we shall speak of $\eps$ as the \emph{ellipticity} of the disk's motion.\\

\subsection{Hamiltonian formulation - Newtonian level}

After this short review we shall now derive the Hamiltonian
formulation of the disk's dynamics. On the Newtonian level this can be
easily achieved by exploiting the analogy of (\ref{disk8}) to the
radial equation of the Keplerian problem. The Lagrangian, leading to
Eq. (\ref{disk8}), reads
\begin{eqnarray*}
      L=\al\left[\frac{\dot{r}_d^2}{2}+\frac{\beta}{r_d}+\frac{\gamma}{\rd^2}\right],
\end{eqnarray*}
$\alpha, \beta$ and $\gamma$ being coefficients which have to be
determined. The Euler-Lagrangian equation corresponding to this
ansatz reads
\begin{eqnarray*}
       \frac{d}{dt}\frac{\pa L}{\pa\dot{\rd}}-\frac{\pa L}{\pa
       \rd}=0\qquad\Longrightarrow \qquad 
       \ddot{r}_d+\frac{\beta}{\rd^2}+\frac{2\gamma}{\rd^3}=0.
\end{eqnarray*}
Comparing this with Eq. (\ref{disk8}), we have to identify
\begin{eqnarray*}
      \beta=2C,\qquad \gamma=-\frac{h^2}{2}=-CR_d\xi^2.
\end{eqnarray*}
Substituting the canonical conjugate momentum to $\rd$, 
\begin{eqnarray*}
     p_r=\frac{\pa L}{\pa \dot{r}_d}=\alpha\dot{r}_d,
\end{eqnarray*}
the  Newtonian Hamiltonian reads
\begin{eqnarray}
\label{disk12}
      H_{d}^{(N)}=\frac{p_r^2}{2\alpha}-\al\left(\frac{\beta}{\rd}+\frac{\gamma}{\rd^2}\right)=\frac{1}{2\al}\left(p_r^2+\frac{p_\vp^2}{\rd^2}\right)-\frac{2C\al}{\rd},
\end{eqnarray}
where we have identified $p_\vp^2\equiv 2\al^2CR_d\xi^2$. To fix the
value of $\al$ we require that $H_d^{(N)}$ should be the conserved
energy of the disk. The later is given by 
\begin{eqnarray*}
     E_d=\int d^3x \frac{1}{2}\rho(v^2-U)=\frac{3\pi}{20}\frac{GM_d^2}{R_d}(\xi^2-2)=\frac{2}{5}\frac{C}{R_d}M_d(\xi^2-2).
\end{eqnarray*}
In particular, this must hold at $t=0$, where by virtue of
the initial condition $\rd(0)=R_d$ and $\dot{r}_d(0)=0$, the momentum
conjugate has to fullfil $p_r(0)=0$. Inserting this into Eq. (\ref{disk12}), we end up with
\begin{eqnarray}
\label{disk13}
    H_{d}^{(N)}=\frac{1}{2\al}\left(p_r^2+\frac{p_\vp^2}{\rd^2}\right)-\frac{2C\al}{\rd},\qquad  \al=\frac{2}{5}M_d.
\end{eqnarray}

\subsection{Radiation reaction part}

It is well known that the leading order part of the gravitational radiation is
of the quadrupolar type. In the case of continuous matter  the related
radiation reaction Hamiltonian is given by \cite{Schafer:1990}
\begin{align}
\label{reac1}
     H_{reac}(t)&=\frac{2G}{5c^5}\dddot{Q}_{ij}(t)\int d^3x
     (\frac{\pi_i\pi_j}{\rho_\ast}+\frac{1}{4\pi
     G}\pa_i U_\ast\pa_j U_\ast), 
\end{align}
where
\begin{align}
      Q_{ij}:=\int d^3x \rho(x^ix^j-\frac{1}{3}\delta^{ij}r^2)
\end{align}
is the Newtonian mass-quadrupole tensor of the system.
At Newtonian order we may identify $\rho_\ast=\rho, \pi_i=v^i\rho$,
and $U_\ast=U$.
Using the continuity equation for the mass and the equations of motion
for the matter, the following equation, relating an integral of
non-compact support to an integral of compact support, can be shown to
hold, 
\begin{align}
\label{reac2}
     \int d^3x (\rho v^iv^j+\frac{1}{4\pi
     G}\pa_i U\pa_j U) = \frac{1}{2}\ddot{Q}_{ij}.
\end{align}
Thus we are allowed to write the reaction Hamiltonian of our disk as
\begin{align}
\label{reac3}
      H_{reac}^{(d)}(t)=\frac{G}{5c^5}\dddot{Q}^{(d)}_{ij}(t)\ddot{Q}_{ij}^{(d)}(p,q),
\end{align}
where $\ddot{Q}_{ij}^{(d)}(p,q)$ denotes that $\ddot{Q}_{ij}^{(d)}$
has to be treated as a function of position and momentum variables, i.e.
making use of the equations of motion the second time derivatives have
to be eliminated.
The mass quadrupole tensor of the disk is diagonal,
$Q_{ij}^{(d)}=(Q_{11}^{(d)},Q_{11}^{(d)},-2Q_{11}^{(d)}).$ This
simplifies the Eq. (\ref{reac3}) enormeously. In the appendix the disk's
quadrupole tensor and it's time derivatives are given explicitly. In
particular, we get
\begin{eqnarray}
\label{quadru2}
     \ddot{Q}_{11}^{(d)}(p,q)=\frac{1}{3}
     \left[\frac{1}{\al}(p_r^2+\frac{p_\vp^2}{\rd^2})-\frac{2C\al}{\rd}\right].
\end{eqnarray}
Inserting this into Eq. (\ref{reac3}) we finally arrive at
\begin{align}
\label{reac4}
     H_{reac}^{(d)}(t)=\frac{2G}{5c^5}\dddot{Q}_{xx}^{(d)}(t)\left[\frac{1}{\al}(p_r^2+\frac{p_\vp^2}{\rd^2})-\frac{2C\al}{\rd}\right].
\end{align}

\subsection{Equations of motion}

The total Newtonian plus leading order radiation reaction
(non-autonomous) Hamiltonian of the oscillating rotating disk reads
\begin{align}
    H_{total}^{(d)}=H_d^{(N)}+H_{reac}^{(d)}(t),
\end{align}
where $H_d^{(N)}$ and $H_{reac}^{(d)}$ are given by (\ref{disk13}) and
(\ref{reac4}), respectively. The Hamiltonian equations of motion are
obtained in the usual way as
\begin{align*}
     \dot{p}_i=-\frac{\pa H_{total}^{(d)}}{\pa q^i},\qquad
     \dot{q}_i= \frac{\pa H_{total}^{(d)}}{\pa p_i}.
\end{align*}
However, some care is needed when calculating the reaction part of
the Hamiltonian equations of motion. It is obtained by taking the
derivatives of  $H_{reac}^{(d)}$ with respect to the
canonical conjugate variables, but one \emph{must} take
$\dddot{Q}_{ij}^{(d)}(t)$ as a function of $t$. Only \emph{after}
the differentiation we shall express the third time derivative of the
quadrupole tensor as a function of $p_r$ and $\rd$. Inserting Eq. 
(\ref{quadru2}) the reaction part of the Hamiltonian equations of
motion reads
\begin{align*}
     (\dot{r}_d)_{reac}&=\frac{\pa H_{reac}^{(d)}}{\pa
     p_r}=-\frac{8}{15}\frac{GC}{c^5}\frac{p_r^2}{\al\rd^2},\\
     (\dot{\vp})_{reac}&=-\frac{8}{15}\frac{GC}{\al
     c^5}\frac{p_rp_\vp}{\rd^4}, \\
     (\dot{p}_r)_{reac}&=-\frac{8}{15}\frac{GC}{c^5}\frac{p_r}{\rd^2}\left(\frac{p_\vp^2}{\al \rd^3}-\frac{C\al}{\rd^2}\right),\\
     (\dot{p}_\vp)_{reac}&=0.
\end{align*}
The Hamiltonian equations for the oscillating disk, including leading
order gravitational radiation reaction, read
\begin{align}
\label{reacdisk1}
      \dot{\rd}&=\frac{p_r}{\al}\left[1-\frac{8}{15}\frac{GC}{c^5}\frac{p_r}{\rd^2}\right] \\
\label{reacdisk2}
      \dot{\vp}&=\frac{p_\vp}{\al\rd^2}\left[1-\frac{8}{15}\frac{GC}{c^5}\frac{p_r}{\rd^2}\right] \\
\label{reacdisk3}
      \dot{p}_r&=\frac{p_\vp^2}{\al\rd^3}-\frac{2\al
      C}{\rd^2}-\frac{8}{15}\frac{GC}{c^5}\frac{p_r}{\rd^2}\left(\frac{p_\vp^2}{\al\rd^3}-\frac{C\al}{\rd^2}\right)\\
\label{reacdisk4}
      \dot{p}_\vp&=0.
\end{align}
Note that we do not treat gravitational damping as a secular
effect. The leading order radiation reaction terms enter directly into
the equations of motion.
It is remarkable that, although the total energy of the disk is not conserved,
the angular momentum \emph{is} a conserved quantity at 2.5pN as
indicated by Eq. (\ref{reacdisk4}). This is due to the
rotational symmetry of the disk and well expected. For a binary point-mass system,
however, 2.5pN radiation reaction destroys orbital momentum
conservation, as we shall see in the next section.

\begin{figure}
\begin{center}
\psfrag{d2I20dt}{$\quad \ddot{I}^{\,20}_{disk}$ [$\al c^2$]}
\psfrag{-0.1}{$\ -0.1$}
\psfrag{500}{$500$}
\psfrag{1000}{$1000$}
\psfrag{1500}{$1500$}
\psfrag{2000}{$2000$}
\psfrag{2500}{$2500$}
\psfrag{3000}{$3000$}
\psfrag{0}{$0$}
\psfrag{3500}{$3500$}
\psfrag{-0.2}{$\ -0.2$}
\psfrag{0.1}{$0.1$}
\psfrag{0.2}{$0.2$}
\psfrag{Time}{$t$ [$G\al/c^3$]}
\includegraphics[width=10cm]{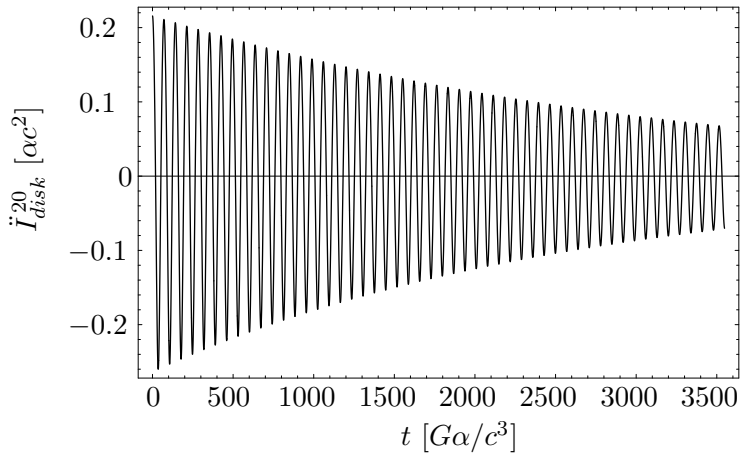}
\caption{In leading order approximation $\ddot{I}^{20}_{disk}$ is the
only nonvanishing contribution to the gravitational radiation field of
the rotating, oscillating disk. In the figure $\ddot{I}^{20}_{disk}$
is plotted for $\xi^2=0.9, R_d(0)=10$ (in units of $G\alpha/c^2$).}
\label{fig:1}
\end{center}
\end{figure}

\subsection{Gravitational waves from oscillating rotating disks}

The 2.5pN radiation reaction terms in
Eqs. (\ref{reacdisk1})-(\ref{reacdisk4}) are the leading order
dissipative terms. Energy is dragged from the oscillating disk by
emission of gravitational waves.\\
Asymptotically, the gravitational radiation field can be represented
by \cite{Thorne:1980ru}
\begin{align}
    h_{ij}^{rad}&=\frac{G}{c^4D}\sum_{l=2}^\infty \sum_{m=-l}^l\left[
    \left(\frac{1}{c}\right)^{l-2}\,
    ^{(l)}I^{lm}(t-D/c)T_{ij}^{E2,lm}(\theta,\phi)\right.\nonumber \\
    &\hspace{2.5cm}\left. +\left(\frac{1}{c}\right)^{l-1}\, ^{(l)}S^{lm}(t-D/c)T_{ij}^{B2,lm}(\theta,\phi)\right],
\end{align}
where $D$ denotes the source-observer distance, the indices $i,j$ refer
to Cartesian coordinates in the asymptotic space, $T_{ij}^{B2,lm}$ and
$T_{ij}^{E2,lm}$ are the pure-spin tensor-spherical harmonics of
magnetic and electric type and $I^{lm}$ and $S^{lm}$ are the spherical
radiative mass and current multipole moments, respectively. The upper pre-index
$(l)$ denotes the number of time derivatives. In leading order
approximation only  $l=2$ terms contribute to the radiation field,
i.e.
\begin{align}
\label{rad}
      h_{ij}^{rad}=\frac{G}{c^4D}\sum_{m=-2}^2 \ddot{I}^{2m} T_{ij}^{E2,2m}(\theta,\phi).
\end{align}
This is  the famous quadrupole radiation.\\
In many cases it is more convenient to work with STF mass multipole
moments $Q_{A_l}$. They are related to the spherically radiative mass multipole
$I^{ij}$  according to, e.g. \cite{Junker:1992}
\begin{align}
      I^{lm}(t)=\frac{16\pi}{(2l+1)!!}\left[\frac{(l+1)(l+2)}{2l(l-1)}\right]^{1/2}Q_{A_l}Y_{A_l}^{lm\ast},
\end{align}
where
\begin{align}
     Y_{A_l}^{lm}:=(-1)^m(2l-1)!!
     \left[\frac{2l+1}{4\pi(l-m)!(l+m)!}\right]^{1/2}
     (\delta^1_{\langle i_1}+i\delta^2_{\langle i_1})\cdots(\delta^1_{i_m}+i\delta^2_{i_m})\delta^3_{i_{m+1}}\cdots \delta^3_{i_l\rangle}.
\end{align}
The brackets denote the symmetric part. In
particular, we get
\begin{align}
    Y^{20}_{i_1i_2}&=3\sqrt{\frac{5}{16\pi}}\delta_{i_1}^3\delta_{i_2}^3,
    \\
    Y^{21}_{i_1i_2}&=-\sqrt{\frac{15}{8\pi}}(\delta_{
    \langle i_1}^1\delta^3_{i_2\rangle}+i\delta^2_{\langle
    i_1}\delta^3_{i_2\rangle}),\\
    Y^{22}_{i_1i_2}&=\sqrt{\frac{15}{32\pi}}(
    \delta_{i_1}^1\delta^1_{i_2}-\delta^2_{i_1}\delta^2_{i_2}+2i\delta_{\langle i_1}^1\delta_{i_2\rangle}^2).
\end{align}
Inserting these expressions, we readily find
\begin{align}
\label{I20}
      I^{20}&=\frac{16\pi}{15}\sqrt{3}Y^{20\ast}_{i_1i_2}Q_{i_1i_2}=4\sqrt{\frac{3\pi}{5}}Q_{33},\\
\label{I21}
      I^{21}&=\sqrt{3}\frac{16\pi}{15}Y^{21\ast}_{i_1i_2}Q_{i_1i_2}=-4\sqrt{\frac{2\pi}{5}}(Q_{13}-iQ_{23}),\\
\label{I22}
      I^{22}&=\sqrt{3}\frac{16\pi}{15}\sqrt{\frac{15}{32\pi}}(Q_{11}-Q_{22}-2iQ_{12}) =2\sqrt{\frac{2\pi}{5}}(Q_{11}-Q_{22}-2iQ_{12}).
\end{align}
The quadrupole tensor of the disk is diagonal and, moreover,
$Q_{11}^{(d)}=Q_{22}^{(d)}$. Hence only 
$\ddot{I}^{\,20}_{disk}$ contributes to the leading order
gravitational radiation field of the disk. In terms of $r_d,p_r, \vp,
p_\vp$ the second time derivative of $Q_{33}^{(d)}$ reads
\begin{eqnarray*}
     \ddot{Q}_{33}^{(d)}=-\frac{2}{3}\left[\frac{1}{\al}(p_r^2+\frac{p_\vp^2}{\rd^2})-\frac{2C\al}{\rd}\right].
\end{eqnarray*}
Inserting this into Eq. (\ref{I20}), the leading order time dependence
of the gravitational waveform
of the oscillating rotating disk of dust is given by
\begin{align}
\label{I20disk1}
     \ddot{I}^{20}_{disk}=-8\sqrt{\frac{\pi}{15}}\left[\frac{1}{\al}(p_r^2+\frac{p_\vp^2}{\rd^2})-\frac{2C\al}{\rd}\right].
\end{align}
In figure \ref{fig:1} the $\ddot{I}^{20}$ component of the leading
order gravitational radiation field is given for a particular example.\\ 
Exploiting the analogy of the analytic solution Eq. (\ref{disk11}) we
may express $\ddot{I}^{20}_{disk}$ in terms of $u$. To that end we
start with Eq. (\ref{quadru2}), where $\ddot{Q}_{11}^{(d)}$ is given
as a function of the Newtonian energy $E_d$ and $\rd$. Inserting
\begin{align*}
     E_d=-\frac{\al C}{R_d}(1+\eps),\qquad
     \rd=a_d(1-\eps\cos u)=\frac{R_d}{1+\eps}(1-\eps\cos\, u)
\end{align*}
into (\ref{quadru2}) we readily obtain
\begin{align*}
    \ddot{Q}_{11}^{(d)}=\frac{2\al}{3}\left(\frac{E_d}{\al}+\frac{C}{\rd}\right) =\frac{2}{3}E_d\left(1-\frac{1}{1-\eps\cos u}\right),
\end{align*}
and hence
\begin{align*}
     \ddot{I}^{20}_{disk}=4\sqrt{\frac{3\pi}{5}}
     \ddot{Q}_{33}^{(d)}=-16\sqrt{\frac{\pi}{15}}
     E_d\left(1-\frac{1}{1-\eps\cos\ u}\right).
\end{align*}
This expression is quite similar to the result given for  $\ddot{I}^{20}_{orb}$
(see Eq. (\ref{I20ana})).

\section{The orbital motion reviewed}

\subsection{Hamiltonian formalism}

The dynamics of a binary point-mass system, including leading order
gravitational radiation reaction, is well known, and we shall only
briefly review the basic points. The orbital  Hamiltonian reads
\begin{align}
      H_{orb}=H_{orb}^{(N)}+H^{(o)}_{reac}(t).
\end{align}
Calculating the leading order reaction Hamiltonian according to Eq.
(\ref{reac1}) we arrive at~\cite{Kokkotas:1995xe}
\begin{align}
\label{reac1orb}
     H_{reac}^{(o)}=\frac{2}{5}\frac{G}{c^5}\dddot{Q}_{ij}^{(o)}\left[\frac{P_iP_j}{\mu}-G\M\mu\frac{R^iR^j}{R^3}\right],
\end{align}
and the Newtonian Hamiltonian is given by
\begin{align}
      H_{orb}^{(N)}=\frac{1}{2\mu}\left(P_R^2+\frac{P_\Phi^2}{R^2}\right)-\frac{G\M\mu}{R},
\end{align}
where $\M$ is the total and $\mu$ the reduced mass, respectively.
The time evolution of the system is governed by the Hamiltonian
equations of motion,
\begin{eqnarray*}
    \dot{P}_i=-\frac{\pa H_{orb}}{\pa R^i},\qquad
    \dot{R}^i=\frac{\pa H_{orb}}{\pa P_i}.
\end{eqnarray*}
Note that $\dddot{Q}_{ij}^{(o)}(t)$ has to be taken as a function of
time. Only after the differentiation it shall be expressed as a function
of $P_R,P_\Phi,R$ and $\Phi$. In polar coordinates, the Hamiltonian
equations of the orbital motion read
\begin{align}
\label{eomorbit}
    \dot{R}&=\frac{P_R}{\mu}-\frac{8}{15}\frac{G^2}{R^2\nu
    c^5}\left[2P_R^2+6\frac{P_\Phi^2}{R^2}\right], \nonumber\\
    \dot{\Phi}&=\frac{P_\Phi}{\mu R^2}-\frac{8}{3}\frac{G^2}{c^5\nu
    R^4}P_RP_\Phi,\\
    \dot{P}_\Phi&=-\frac{8}{5}\frac{G^2 P_\Phi}{c^5\nu
    R^3}\left[\frac{2G\M^3\nu^2}{R}+\frac{2P_\Phi^2}{R^2}-P_R^2\right], \nonumber\\
    \dot{P}_R&=\frac{P_\Phi^2}{\mu
    R^3}-\frac{G\M\mu}{R^2}+\frac{8}{3}\frac{G^2P_R}{c^5R^4}\left[\frac{G\M^3\nu}{5}-\frac{P_\Phi^2}{\nu R}\right]
\end{align}
where $\nu=\mu/\M$.\\
For numerical calculations we shall apply rescaled variables.
 Defining $X,P_R',P_\Phi'$ and $\tau$ by
\begin{align}
\label{scaling}
       R=:\frac{G\M}{c^2}X,\qquad
       P_R=:\mu cP_R',\qquad
       P_\Phi=:\frac{G\M\mu}{c}P_\Phi',\qquad
       t=:\frac{G\M}{c^3}\tau
\end{align}
we arrive at
\begin{align}
\label{scaledorbit1}
      \frac{dX}{d\tau}&=P_R'-\frac{8}{15}\frac{\nu}{X^2}
                        \left[2P_R'^2+6\frac{P_\Phi'^2}{X^2}\right] \\
      \frac{d\Phi}{d\tau}&=\frac{P_\Phi'}{X^2}-
                          \frac{8}{3}\frac{\nu}{X^4}P_R'P_\Phi' ,\\
      \frac{dP'_\Phi}{d\tau}&=-\frac{8}{5}\frac{\nu P_\Phi'}{X^3}
                  \left[\frac{2}{X}+\frac{2P_\Phi'^2}{X^2}-P_R'^2\right] ,\\
      \frac{P'_R}{d\tau}&=\frac{P_\Phi'^2}{X^3}-\frac{1}{X^2}+
                          \frac{8}{3}\frac{\nu P_R'}{X^4}
                          \left[\frac{1}{5}-\frac{P_\Phi'^2}{X}\right].
\end{align}

\subsection{Leading order gravitational waveforms}

The leading order gravitational radiation field is given in Eq.
(\ref{rad}). If we identify the orbital plane with the $\Theta=\pi/2$
- plane, $Q_{13}^{(o)}$ and $Q_{23}^{(o)}$ vanish and  thus only
$I^{20}_{orb}$ and $I^{22}_{orb}$ are
nonzero. The time derivatives of  $Q_{ij}^{(o)}$ as functions of the
canonical variables are given in the Appendix B. Inserting this in
(\ref{I22}) and (\ref{I20}), respectively, we find
\begin{align}
\label{I22orbtotal}
     \ddot{I}^{22}_{orb}&=4\sqrt{\frac{2\pi}{5}}e^{-2i\Phi}\left[\frac{P_R^2}{\mu}-\frac{P_\Phi^2}{\mu R^2}-\frac{G\M\mu}{R}-2i\frac{P_RP_\Phi}{\mu R}\right] \nonumber \\
    &= 4\sqrt{\frac{2\pi}{5}}e^{-2i\Phi}\mu c^2\left[P_R'^2-\frac{P_\Phi'^2}{X^2}-\frac{1}{X}-2i\frac{P_R'P_\Phi'}{X}\right]
\end{align}
or, if splitted in real and imaginary parts,
\begin{align}
      \Re(\ddot{I}^{22}_{orb})&=4\sqrt{\frac{2\pi}{5}}\mu c^2
      \left[\cos(2\Phi)
      \left\{P_R'^2-\frac{P_\Phi'^2}{X^2}-\frac{1}{X}\right\}
      -2\sin(2\Phi)\frac{P_R'P_\Phi'}{X}\right]\\
      \Im(\ddot{I}^{22}_{orb})&=
      -4\sqrt{\frac{2\pi}{5}}\mu c^2
      \left[\sin(2\Phi)
      \left\{P_R'^2-\frac{P_\Phi'^2}{X^2}-\frac{1}{X}\right\}
      +2\cos(2\Phi)\frac{P_R'P_\Phi'}{X}\right],
\end{align}
and
\begin{align}
\label{I20orb}
      \ddot{I}^{20}_{orb}=-8\sqrt{\frac{\pi}{15}}\mu c^2\left[P_R'^2+\frac{P_\Phi'^2}{X^2}-\frac{1}{X}\right].
\end{align}

\subsection{Gravitational waveforms -- analytical results}

The contributions to the gravitational waveform in Eqs. (\ref{I22orbtotal}) and (\ref{I20orb}) are given in terms of $P_R',P_{\Phi}', \Phi$
and $X$ . These expressions are applied when solving the dynamics
fully numerically. It is instructive, however, to compare those results
with semi-analytical ones where the conservative dynamics is solved
fully analytically and where the gravitational radiation damping is
treated as a secular effect.
If the two components are well separated, the curves turn out to be
in good agreement.  At distances of around 10 $G\M/c^2$,
$\ddot{I}^{22}$ as calculated by the secular approach is in a good
agreement to the one calculated by full numerical simulation over a few
periods only. \\
The  parametrized solution to the Newtonian problem reads
\begin{eqnarray*}
    X=a'(1-e\,\cos u),\qquad
    \frac{2\pi}{P}t\equiv n\cdot t=u-e\cos u.
\end{eqnarray*}
Here $u$ is the mean anomaly, the period $P$ is  given by
\begin{align}
\label{kepler2}
     n=\frac{2\pi}{P}=\sqrt{\frac{G\M}{a^{'3}}},
\end{align}
and the azimuthal angle $\Phi$ is related to $u$ by
\begin{align}
\label{kepler3}
      \Phi=2\arctan\left[\sqrt{\frac{1+e}{1-e}}\tan\frac{u}{2}\right].
\end{align}
From Eq. (\ref{kepler3}) we readily obtain
\begin{align}
\label{kepler4}
     \dot{\Phi}=\frac{\sqrt{1-e^2}}{1-e\,\cos
     u}\,\dot{u}=\frac{n\sqrt{1-e^2}}{(1-e\,\cos u)^2}.
\end{align}
The Newtonian energy of the orbit is $E=\mu c^2 E'$ with
$E'=-\frac{1}{2a'}$. Using this relation, it is now easy to derive from
the eqs. (\ref{I20}) and (\ref{I22}) 
the leading order contributions \cite{Junker:1992}
\begin{align}
\label{I22ana}
    \ddot{I}^{22}&=8\sqrt{\frac{2\pi}{5}} e^{-2i\Phi}\mu
    c^2E'\left[
    1-\frac{1}{A(u)}+\frac{2(1-e^2)}{A(u)^2}+2i\frac{e\sqrt{1-e^2}\sin u}{A(u)^2}\right],\\
\label{I20ana}
    \ddot{I}^{20}&=-16\sqrt{\frac{\pi}{15}}\mu c^2 E'\left[1-\frac{1}{A(u)}\right],    
\end{align}
where $A(u)=1-e\cos u$. In particular, $\ddot{I}^{20}$ vanishes
identically for circular orbits. The luminosity of the gravitational
wave emission in leading order approximation is given by
\begin{align*}
    \mathcal{L}=\frac{G}{5c^5}\frac{d^3Q_{ij}}{dt^3}\frac{d^3Q_{ij}}{dt^3}.
\end{align*}
Averaging $\mathcal{L}$ over the orbital period gives the leading
order energy loss
\begin{align*}
     \left\langle\frac{dE}{dt}\right\rangle&=-\frac{1}{P}\int_0^P \mathcal{L}\,dt=
     -\frac{32}{5c^5}\frac{G^4\M^3\mu^2}{a^5(1-e^2)^{7/2}}
     \left[1+\frac{73}{24} e^2+\frac{37}{96} e^4\right].
\end{align*}
while the decay of the orbital eccentricity is given by
\begin{align*}
      \left\langle \frac{de}{dt}\right\rangle
      =-\frac{1}{15}\frac{\nu}{G\M c^5}\left(\frac{G\M}{a}\right)^4
      \frac{e}{(1-e^2)^{5/2}}\,(304 +121 e^2).
\end{align*}

\section{The binary system}

In previous chapters we have considered the Newtonian oscillating
rotating disk and the orbital motion separately. In the following we
shall focus on a binary system composed of a point-like object
and a dusty disk. Due to tidal
interaction neither the internal dynamics of the
disk nor the orbital motion are completely
independent. In fact, the influence of tidal coupling on the form
of the orbit and on the inspiral process can be rather strong, as we
will see later on. For simplicity we assume the orbital plane to be
the $z=0$ plane, i.e. the plane of the
disk. 

\subsection{Tidal coupling on the Newtonian level}

The Newtonian interaction potential for a general matter distribution reads 
\begin{align}
      U&=-\frac{G}{2}\iint\frac{\rho(\vec r)\rho(\tilde{\vec r})}{|\vec
      r-\tilde{\vec r}|} \,d^3rd^3\tilde{r}\nonumber \\
      &=-\frac{G}{2}\iint\frac{(\rho_p(\vec r)+\rho_d(\vec
      r))(\rho_p(\tilde{\vec r})+\rho_d(\tilde{\vec r})}{|\vec r
      -\tilde{\vec r}|}\,d^3r d^3\tilde{r}.
\end{align}
Not taking into account self-interaction terms, we are left with
\begin{align}
\label{tidal1}
     U_{int}=U_{int}^{(1,2)}+U_{int}^{(2,1)}=-G\int\frac{\rho_d(\vec
     r)\rho_p(\tilde{\vec r})}{|\vec r-\tilde{\vec r}|}\,d^3
     rd^3\tilde{r}.
\end{align}
One of the two volume integrals is trivial since the point mass density is
given by $\rho_p(\tilde{\vec r})=M_p\delta(\tilde{\vec r}-\vec
R)$. The remaining integral reads
\begin{align}
\label{tidal2}
    U_{int}=-GM_p\int\frac{\rho_d(\vec r)}{|\vec r-\vec R|}\, d^3r.
\end{align}
Inserting the density of the disk $\rho_d=\delta(z)\Sigma(r,t)$,
where the surface density $\Sigma$ is defined by Eq. (\ref{disk3})
and where the origin of the coordinate system has been shifted to the
center of the disk, we end up with
\begin{align}
     U_{int}&=-GM_p\sigma(t)\int_0^{2\pi}d\vp\int_0^{r_d}dr r
     \frac{\sqrt{1-r^2/r_d^2}}{\sqrt{r^2+R^2-2rR\cos\vp}} \nonumber
     \\
\label{tidalhelp}     
    &=-GM_p\frac{\sigma(t)}{R}\, J,
\end{align}
where  $J$ is defined as
\begin{align}
\label{tidal3}
    J:=\int_0^{2\pi}d\vp\int_0^{r_d}dr\,r\frac{\sqrt{1-(r/r_d)^2}}{\sqrt{1+(r/R)^2-2(r/R)\cos\vp}}.
\end{align}
Unfortunately, this integral seems not to be solvable
analytically for $R>\rd$ (in case of $R<\rd$, see e.g. Kley and
Sch\"afer \cite{Schafer:1994a}). Expanding the integrand in terms of
Legendre polynoms, we obtain
\begin{align}
\label{tidalhelp2}
     J&=\int_0^{2\pi}d\vp\int_0^{r_d}dr\,r\sqrt{1-\left(\frac{r}{r_d}\right)^2}\sum_{l=0}^\infty \left(\frac{r}{R}\right)^l P_l(\cos\vp) \nonumber\\
      &=r_d^2\sum_{l=0}^\infty
    \left(\frac{r_d}{R}\right)^l\int_0^{2\pi}d\vp
    P_l(\cos\vp)\int_0^1\sqrt{1-x^2}x^{l+1} dx.  
\end{align}
Defining
\begin{align*}
     Q_l:=\int_0^1x^{l+1}\sqrt{1-x^2} dx =\frac{\sqrt{\pi}}{4}\frac{\Gamma(1+\frac{l}{2})}{\Gamma(\frac{5+l}{2})},
\end{align*}
we end up with
\begin{align*}
     J=r_d^2\sum_{l=0}^\infty \left(\frac{r_d}{R}\right)^lQ_l \int_0^{2\pi}P_l(\cos\vp')d\vp'.
\end{align*}
The integral $\int_0^{2\pi}P_{l}(\cos\vp)d\vp$ is only
nontrivial for even numbers of $l$. With the substitutions
$y=\cos\vp,\ d\vp=-\frac{dy}{\sqrt{1-y^2}}$ the case $l=2n$ can be rewritten
as 
\begin{align*}
    \int_0^{2\pi}P_{2n}(\cos\vp')d\vp' =2\int_{-1}^1\frac{P_{2n}(y)}{\sqrt{1-y^2}}\,dy.
\end{align*}
This is a special case of the more general relation \cite{Gradsteyn}
\begin{align*}
     &\int_{-1}^1(1-y^2)^{\lambda-1}P_\nu^{\mu}(y)dy =\frac{2^\mu
     \pi\Gamma(\lambda+\frac{\mu}{2})\Gamma(\lambda-\frac{\mu}{2})}
     {\Gamma(\lambda+\frac{\nu+1}{2})\Gamma(\lambda-\frac{\nu}{2})
     \Gamma(1+\frac{\nu-\mu}{2})\Gamma(\frac{1-\mu-\nu}{2})},
\end{align*}
which holds for $\Re\,\lambda>\frac{1}{2}|\Re \mu|$. Taking
$\lambda=\frac{1}{2},\ \mu=0,\ \nu=2n$ we arrive at 
\begin{align*}
    \int_{-1}^1\frac{P_{2n}(y)}{\sqrt{1-y^2}}\,dy
    =\frac{\pi\Gamma(\frac{1}{2})^2}{\Gamma(1+n)^2\Gamma(\frac{1}{2}-n)^2} =\left(\frac{\pi}{n!\Gamma(\frac{1}{2}-n)}\right)^2,
\end{align*}
and thus
\begin{align*}
     \int_0^{2\pi}P_{2n}(\cos\vp)d\vp=2\left(
      \frac{\pi}{n!\Gamma(\frac{1}{2}-n)}\right)^2.
\end{align*}
Inserting this into Eq. (\ref{tidalhelp2}) the tidal potential
Eq. (\ref{tidalhelp}) reads
\begin{align}
\label{tidalpot1}
     U_{int}&=-\frac{GM_p}{2R}\sigma(t)r_d^2\pi\sum_{l=0}^\infty
     \left(\frac{r_d}{R}\right)^{2l}W_l=-\frac{3}{4}\frac{GM_pM_d}{R}\sum_{l=0}^{\infty}\left(\frac{\rd}{R}\right)^l W_l  \\
     &=U_{orb}+U_{tidal} ,\nonumber
\end{align}
where
\begin{align}
\label{Wl}
    W_l:=\frac{\pi^{3/2}}{\left[\left(\frac{l}{2}\right)!\Gamma\left(\frac{1-l}{2}\right)^2\Gamma\left(\frac{5+l}{2}\right)\right]}, \qquad l\ \text{even},
\end{align}
and $W_l=0$ for odd values of $l$. 
It is easy to see that the $l=0$ term in Eq. (\ref{tidalpot1}) gives just the point-mass interaction,
\begin{align}
     U_0=-\frac{3}{4}\frac{GM_pM_d}{R}\frac{\pi^{3/2}}{\Gamma(\frac{1}{2})^2\Gamma(\frac{5}{2})}=-\frac{GM_pM_d}{R}\equiv U_{orb},
\end{align}
where we used  $\Gamma(\frac{1}{2})=\sqrt{\pi}$ and
$\Gamma(\frac{5}{2})=\frac{3}{4}\sqrt{\pi}$.
The tidal interaction potential can be immediately read off
from Eq. (\ref{tidalpot1}),
\begin{align}
\label{tidal4}
     U_{tidal}=& -\frac{3}{4}\frac{G M_pM_d}{R}\sum_{l=2}^\infty
     \left(\frac{\rd}{R}\right)^l W_l.
\end{align}
As expected, the first nonvanishing contribution comes from the
quadrupole term \cite{Mora:2003wt}.
Correspondingly, we expect the tidal interaction to
introduce a periastron shift into the Newtonian orbit (see e.g. \cite{Landau}).

\subsection{Radiation reaction Hamiltonian and Hamiltonian equations
of motion}

Though it is tempting to take the leading order radiation reaction
Hamiltonian of the binary system as the superposition of the
radiation reaction Hamiltonians of the isolated disk and pure orbital
motion, respectively, a careful analysis shows that coupling
between orbital and disk quadrupole moment already enters at leading
order. This can be easily  seen from the general 2.5 pN reaction
Hamiltonian of a fluid configuration \cite{Schafer:1990},
\begin{align}
\label{reac5}
     H_{reac}&=\frac{2G}{5c^5}\dddot{Q}_{ij}(t)\int d^3r\left(\rho
     v^iv^j+\frac{1}{4\pi G} \pa_iU\pa_jU\right)\nonumber \\
     &=\frac{G}{5c^5}\left(\dddot{Q}_{ij}^{(o)}(t)+\dddot{Q}_{ij}^{(d)}(t)\right)\left(\ddot{Q}_{ij}^{(o)}(R^a,P_a)+\ddot{Q}_{ij}^{(d)}(x^a,p_a)\right)
\end{align}
Here $Q_{ij}^{(d)}$ and $Q_{ij}^{(o)}$ are the STF mass quadrupole
moments of the disk and the orbit, respectively, $R^i,P_i$ denote
the canonical conjugate variables of the orbital
motion and $x^i,p_i$ are the canonical conjugate variables of the disk
relative to the disk's center. With $H_{reac}^{(d)}$ and
$H_{reac}^{(N)}$ given by the eqs. (\ref{reac4}) and
(\ref{reac1orb}) the leading order reaction Hamiltonian of
the binary system is 
\begin{align}
      H_{reac}(t)&=H_{reac}^{(d)}(t)+H_{reac}^{(o)}(t) +\frac{G}{5c^5}\left[\dddot{Q}_{ij}^{(d)}(t)\ddot{Q}_{ij}^{(o)}(R^i,P_i)+\dddot{Q}_{ij}^{(o)}(t)\ddot{Q}_{ij}^{(d)}(x^i,p_i)\right].
\end{align}
The last two terms describe the coupling between disk and orbital
quadrupole moments. Finally, using the diagonal structure of the
disk's mass quadrupole tensor we arrive at
\begin{align}
\label{reac6}
       H_{reac}(t)&=H_{reac}^{(d)}(t)+H_{reac}^{(o)}(t)+
       \frac{G}{5c^5}
       \left[\dddot{Q}_{11}^{(o)}(t)+\dddot{Q}_{22}^{(o)}(t) -2\dddot{Q}_{33}^{(o)}(t)
       \right]
       \ddot{Q}_{11}^{(d)}(x^i,p_i) \nonumber \\
       &+\frac{G}{5c^5}\dddot{Q}_{11}^{(d)}(t)\left[
       \ddot{Q}_{11}^{(o)}(R^i,P_i)+\ddot{Q}_{22}^{(o)}(R^i,P_i)-2\ddot{Q}_{33}^{(o)}(R^i,P_i)\right] \nonumber \\
       &=H_{reac}^{(d)}+H_{reac}^{(o)}(t)- \frac{3G}{5c^5}\left[\dddot{Q}_{33}^{(o)}(t)\ddot{Q}_{11}^{(d)}(x^i,p_i)+\dddot{Q}_{11}^{(d)}(t)\ddot{Q}_{33}^{(o)}(R^i,P_i)\right].
\end{align}
As before, the radiation reaction parts of the Hamiltonian equations
of motion are obtained by differentiating $H_{reac}$ with respect to
the generalized coordinates and momenta. Thus the reaction part of the
equations of motion reads,
\begin{align}
\label{reactotal1}
     (\dot{\rd})_{reac}&=-\frac{8}{15}\frac{GC}{\al
     c^5}\frac{p_r^2}{\rd^2}
     - \frac{4}{15}\frac{G^2\M}{\al
     c^5}\frac{p_rP_R}{R^2} \\
\label{reactotal2}     
     (\dot{\vp})_{reac}&=-\frac{8}{15}\frac{GC}{\al
     c^5}\frac{p_rp_\vp}{\rd^4} - \frac{4}{15}\frac{G^2\M}{\al
     c^5}\frac{p_\vp P_R}{R^2\rd^2} \\
\label{reactotal3}     
     (\dot{p}_r)_{reac}&=\frac{8}{15}\frac{GC}{c^5}\frac{p_r}{\rd^4}
     \left[-\frac{p_\vp^2}{\al\rd}+ C\al\right] + 
     \frac{4}{15}\frac{G^2\M}{c^5}\frac{P_R}{R^2}
     \left[ \frac{C\al}{\rd^2}-\frac{p_\vp^2}{\al\rd^3} \right] \\
\label{reactotal4}
     (\dot{p}_\vp)_{reac}&=0 \\
\label{reactotal5}     
     (\dot{R})_{reac}&=-\frac{16}{5}\frac{G^2\M}{\mu
     c^5R^2}\left[\frac{P_\Phi^2}{R^2}+\frac{1}{3}P_R^2\right]
     -\frac{8}{15}\frac{GC}{\mu c^5}\frac{p_rP_R}{\rd^2} \\
\label{reactotal6}
     (\dot{\Phi})_{reac}&=-\frac{8}{3}\frac{G^2\M}{\mu
     c^5}\frac{P_RP_\Phi}{R^4}-\frac{8}{15}\frac{GC}{\mu c^5}\frac{p_r
     P_\Phi}{R^2\rd^2} \\
\label{reactotal7}     
     (\dot{P}_R)_{reac}&=-\frac{8}{3}\frac{G^2\M}{c^5}\frac{P_R}{R^4}\left[\frac{P_\Phi^2}{\mu R}-\frac{1}{5}G\M\mu\right]+\frac{4}{15}\frac{GC}{c^5}\frac{p_r}{\rd^2 R^2}\left[G\M\mu -\frac{2P_\Phi^2}{\mu R}\right]\\
\label{reactotal8}     
     (\dot{P}_\Phi)_{reac}&=\frac{8}{5}\frac{G^2\M}{c^5}\frac{P_\Phi}{R^3}\left[\frac{P_R^2}{\mu}-\frac{2G\M\mu}{R}-2\frac{P_\Phi^2}{\mu R^2}\right].
\end{align}
Note that the coupling of disk and orbital mass
quadrupole moments  is already present
at leading order. However, numerically these terms turn out to be very
small even in close orbits and thus can be neglected in most scenarios.

\subsection{Hamiltonian equations of motion}

The total Hamiltonian of the binary takes the form 
\begin{align}
\label{total1}
     H_{tot}=H_{orb}^{(N)}+H_{d}^{(N)}+H_{tidal}+H_{reac},
\end{align}
where
\begin{align}
     H_{orb}^{(N)}&=\frac{1}{2\mu}\left(P_R^2+\frac{P_\Phi^2}{R^2}\right)-\frac{G\M\mu}{R},\\
     H_{d}^{(N)}&=\frac{1}{2\al}\left(p_r^2+\frac{p_\vp^2}{\rd^2}\right)-\frac{2C\al}{\rd}, \\
     H_{tidal}&=-\frac{3}{4}\frac{G\M\mu}{R}\sum_{l=2}^\infty
     \left(\frac{\rd}{R}\right)^l W_l 
\end{align}
and where $H_{reac}$ is given by Eq. (\ref{reac6}). The radiation reactive part
of the equations of motion has been calculated in the previous section
and is given by the Eqs. (\ref{reactotal1}) - (\ref{reactotal8}). The
dynamics of the binary, including tidal coupling and leading order
gravitational radiation reaction effects, is described by the
following set of first order differential equations:
\begin{align}
\label{eom1}
      \dot{P}_R&=\frac{P_\Phi^2}{\mu  R^3} -\frac{G\M\mu}{R^2}
      -\frac{3}{4}\frac{G\M\mu}{R^2}\sum_{l=2}^\infty
      (l+1)\left(\frac{\rd}{R}\right)^l W_l
      -\frac{8}{3}\frac{G^2\M}{\mu c^5}\frac{P_RP_\Phi^2}{R^5}
      \nonumber \\
      &\quad
      +\frac{8}{15}\frac{G^3\M^2\mu}{c^5}\frac{P_R}{R^4}
      +\frac{4}{15}\frac{GC}{c^5}\frac{p_r}{\rd^2R^2}
      \left[G\M\mu-2\frac{P_\Phi^2}{\mu R}\right] \\
\label{eom2}
      \dot{P}_\Phi&=\frac{8}{5}\frac{G^2\M}{c^5}\frac{P_\Phi}{R^3}
      \left[\frac{P_R^2}{\mu}-\frac{2G\M\mu}{R}-
      2\frac{P_\Phi^2}{\mu R^2}\right]\\
\label{eom3}    
      \dot{R}&=\frac{P_R}{\mu}-\frac{16}{5}\frac{G^2\M}{\mu c^5}
      \left[\frac{P_\Phi^2}{R^4}+\frac{1}{3}\frac{P_R^2}{R^2}\right] -
      \frac{8}{15}\frac{GC}{\mu c^5}\frac{p_rP_R}{\rd^2}  \\
\label{eom4}      
      \dot{\Phi}&=\frac{P_\Phi}{\mu R^2}-\frac{8}{3}\frac{G^2\M}{\mu
      c^5}\frac{P_R P_\Phi}{R^4} -\frac{8}{15}\frac{GC}{\mu
      c^5}\frac{p_rP_\Phi}{R^2\rd^2} \\
\label{eom5}      
      \dot{\rd}&=\frac{p_r}{\al}-\frac{8}{15}\frac{GC}{\al
      c^5}\frac{p_r^2}{\rd^2}-\frac{4}{15}\frac{G^2\M}{\al
      c^5}\frac{p_r P_R}{R^2}  \\
\label{eom6}      
      \dot{\vp}&=\frac{p_\vp}{\al \rd^2}-\frac{8}{15}\frac{GC}{\al
      c^5}\frac{p_r p_\vp}{\rd^4} -\frac{4}{15}\frac{G^2\M}{\al
      c^5}\frac{p_\vp P_R}{\rd^2 R^2}\\
\label{eom7}      
      \dot{p}_r&= \frac{p_\vp^2}{\al
      \rd^3}-\frac{2C\al}{\rd^2}+\frac{3}{4}\frac{G\M\mu}{R^2}
      \sum_{l=2}^\infty l \left(\frac{\rd}{R}\right)^{l-1}W_l +
      \frac{8}{15}\frac{GC}{c^5}\frac{p_r}{\rd^4}\left[ C\al -
      \frac{p_\vp^2}{\al \rd}\right] \nonumber \\
      & \quad +\frac{4}{15}\frac{G^2\M}{c^5}\frac{P_R}{\rd^2R^2}
      \left[C\al-\frac{p_\vp^2}{\al \rd}\right] \\
\label{eom8}      
      \dot{p}_\vp&=0.
\end{align}
In order to solve these equations numerically we shall introduce
scaled variables. We choose the same scaling as in
Eq. (\ref{scaling}) for the disk variables
\begin{align}
\label{scaling2}
      p_r=\mu cp_r',\qquad
      p_\vp=\frac{G\M\mu}{c} p_\vp',\qquad
      \rd=\frac{G\M}{c^2} \xd.
\end{align}
Substituting this into the Eqs. (\ref{eom1} - \ref{eom8}) and defining
\begin{align*}
     A:=\frac{C}{G\M}=\frac{3\pi}{8}\frac{M_d}{\M},\qquad
     B:=\frac{\al}{\mu}=\frac{2}{5}\frac{\M}{M_p}
\end{align*}     
we finally arrive at
\begin{align}
\label{scal1}
      \frac{dX}{d\tau}&= P_R'-\frac{16}{5}\frac{\nu}{X^2}
      \left(\frac{P_\Phi'^2}{X^2}
      +\frac{1}{3}P_R'^2\right)
      -\frac{8}{15}\frac{A\nu}{x_d^2}p_r'P_R' \\
\label{scal2}
       \frac{d\Phi}{d\tau}&=\frac{P_\Phi'}{X^2}
       -\frac{8}{3}\nu\frac{P_\Phi' }{X^2}\left[\frac{P_R'}{X^2}+
       \frac{A}{5}\frac{p_r'}{\xd^2}\right] \\
\label{scal3}       
      \frac{dP_\Phi'}{d\tau}&=\frac{8}{5}\nu\frac{P_\Phi'}{X^3}
      \left[P_R'^2-\frac{2}{X}-2\frac{P_\Phi'^2}{X^2}\right] \\
\label{scal4}
      \frac{dP_R'}{d\tau}&=\frac{P_\Phi'^2}{X^3}-\frac{1}{X^2}
      -\frac{3}{4}\frac{1}{X^2}\sum_{l=2}^\infty (l+1)
      \left(\frac{\xd}{X}\right)^l W_l -
      \frac{8}{3}\frac{\nu}{X^5}P_R'P_\Phi'^2 +
      \frac{8}{15}\frac{\nu}{X^4}P_R' \nonumber \\
      & \quad
      + \frac{4}{15}\frac{A\nu}{X^2x_d^2}
      \left[1-2\frac{P_\Phi'^2}{X}\right]p_r' \\
\label{scal5}
      \frac{dp_r'}{d\tau}&=\frac{1}{B}\frac{p_\vp'^2}{\xd^3}-
      2B\frac{A}{\xd^2}+
      \frac{3}{4}\frac{1}{X^2}\sum_{l=2}^\infty l
      \left(\frac{\xd}{X}\right)^{l-1}W_l +
      \frac{8}{15}\frac{A\nu}{x_d^4}
      \left[BA
      -\frac{1}{B}\frac{p_\vp'^2}{\xd}
      \right]p_r' \nonumber \\
      &\quad + 
      \frac{4}{15}\frac{\nu P_R'}{\xd^2X^2}\left[AB
      -\frac{1}{B}\frac{p_\vp'^2}{\xd}\right] \\
\label{scal6}
      \frac{d\vp}{d\tau}&=\frac{1}{B}\left[\frac{p_\vp'}{\xd^2}
      -\frac{8}{15}\frac{A\nu}{x_d^4}p_r'p_\vp'
      -\frac{4}{15}\frac{\nu}{X^2\xd^2}p_\vp'P_R'\right] \\
\label{scal7}      
      \frac{d\xd}{d\tau}&=\frac{1}{B}\left[p_r'
      -\frac{8}{15}A\nu\frac{p_r'^2}{\xd^2}
      -\frac{4}{15}\frac{\nu}{X^2}p_r'P_R' \right]\\
\label{scal8}      
      \frac{d p_\vp'}{d\tau}&=0. 
\end{align}

\subsubsection{Initial conditions}

Both, the dynamics of the binary as well as the leading order
gravitational waveform depend on the choice of initial conditions.
Various initial conditions correspond to various initial phase
differences between orbital and disk angular variables. Let $t=0$ be
the time of the periastron passage where by convention $\Phi(0)=0$.
The initial values for the orbital variables shall read
\begin{align}
\label{init1}
     X(0)=a'(1-e), \qquad
     \Phi(0)=0, \qquad
     P_\Phi'(0)=\sqrt{a'(1-e^2)},\qquad
     P_R'=0.     
\end{align}
To fix the initial conditions for the disk variables, we require
\begin{align}
\label{cond2}
     H'_{d}(0)=\frac{1}{2B}\left(p_r'(0)^2+\frac{p_\vp'(0)^2}{x_d(0)^2}\right)-\frac{2AB}{x_d(0)}\stackrel{!}{=}E_d'(0)=-\frac{AB}{X_d}(1+\eps),
\end{align}
where $\eps$ is the ellipticity of
the disk at $t=0$ and $X_{d}$ is the maximal disk radius. Note that
if $x_d(0)=X_d$ the momentum conjugate $p'_{r}(0)$ vanishes. 
The momentum conjugate to $\vp$ is a constant as can be seen from the
equations of motion. It is given by
\begin{eqnarray*}
     p_\vp^2=2\al^2CR_d\xi^2=\frac{G^2\M^2\mu^2}{c^2}p_\vp'^2
     \qquad\Longrightarrow\qquad
     p_\vp'(0)=B\sqrt{2AX_d(1-\eps)}.
\end{eqnarray*}
The initial value of $p_r'$ for a given $x_d(0)$ can be easily derived
from  Eq. (\ref{cond2}), 
\begin{align}
\label{ini2}
    \xd(0)&=x_{d,0},\qquad
    \vp(0)=0,\qquad
     p_\vp'(0)=B\sqrt{2AX_d(1-\eps)}, \\
     p_r'^2(0)&=2B^2A\left[-\frac{1+\eps}{X_d}+\frac{(\eps-1)X_d}{x_{d,0}^2}+\frac{2}{x_{d,0}}\right].
\end{align}

\section{Discussion}

In the last decade several authors investigated tidal and tidal-resonant
effects in binary systems. Close binary neutron stars were studied by
e.g., Sch\"afer and Kokkotas \cite{Kokkotas:1995xe} and Lai and Ho
\cite{Ho:1998hq}. Recently, resonant excitations of white dwarf oscillations in
a compact binary were investigated by Rathore et al. \cite{Rathore:2004gs}.
In these papers the stellar oscillations were treated
numerically. In our model there is only one oscillation mode, but it
is known analytically. 
Although our model is much simpler than the binary neutron stars of
\cite{Kokkotas:1995xe},\cite{Ho:1998hq} or the white dwarf-compact
object binary of \cite{Rathore:2004gs}, it exhibits all essential
relativistic features of a compact binary system to leading order
approximation. Moreover, we
were able to give the Hamiltonian equations of motion in  analytic form. \\
In their recent paper Rathore et. al. investigated a white
dwarf-compact object system with non-rotating white dwarf.
Gravitational damping was taken into account as a secular effect and
the change of the orbital motion due to tidal interaction was
discussed only qualitatively. The symmetry of our disk-compact object
binary enabled us to study the effects, in
particular the influence of tidal interaction on the actual orbit,
in great detail (see Fig. \ref{fig:2}). Note that while the white
dwarf of Rathore et al. was
nonrotating, rotation is essential in our case to stabilize the dusty
disk. The single eigenmode of the oscillating disk corresponds to the
f-mode of Rathore et al. \\
In our paper we studied the oscillations of an
isolated rotating Newtonian disk of dust in Hamiltonian formalism.
Thin oscillating disks of dust with a diameter of a few hundred
kilometers could eventually form after the merging of two white
dwarfs. The similarity of the disk's equation of motion to the
well-known Keplerian problem is remarkable. Exploiting this analogy
one can easily write down the analytic solution for the disk's
oscillations. This allows to \emph{study the effects of
tidal interaction and it's influence on the actual gravitational
waveform in great detail} (see
Figs.
\ref{fig:3},\ref{fig:4},\ref{fig:6},\ref{fig:7},\ref{fig:8},\ref{fig:9}).
For a suitable choice of parameters
(i.e. the initial major semi-axis not being too small) the evolution of
the system can be followed over hundreds or even thousands of orbital
periods with considerable accuracy (see e.g. Fig. \ref{fig:6}, \ref{fig:9}).
Our model might thus be a good test to run data analysis codes that
intend to extract physical
parameters of the binary system from a given gravitational waveform.\\
A particular feature of the tidally coupled binary system is the so-called
tidal resonance (see Fig. \ref{fig:9}). In Newtonian theory the tidal coupling between the
internal dynamics of a star and the orbital dynamics becomes
particularly strong if one of the star's eigenfrequencies is $n$ times
the orbital frequency. Tidal interaction and in particular  
tidal resonances may speed up the inspiral process by a
considerable amount (see Figs. \ref{fig:2} and \ref{fig:3}). In
fact, for a realistic choice of parameters (i.e. of
$M_d$ and $M_p$) the 2:1 resonance leads to an immediate destruction
of the disk-compact object binary. At this point, the analysis
breaks down. Higher order resonances, starting with the 3:1 one, can
be treated without problems for a suitable choice of parameters.\\
Some remarks should be made on the validity of our model. It relies on
the assumption that the disk is not getting disrupted and that the
change in the disk's density does result from the oscillation of the disk
only. In particular we do not take into
account mass accretion on the compact companion. Thus, even for
tidal resonance, the maximal radius of the disk has to be located far
inside it's Roche lobe.
In other words, the maximal disk radius must be considerably
smaller than the periastron distance between the two objects.\\
In a forthcoming paper we will extend our analysis to a system of two
oscillating disks. In this case there will occur quadrupole-quadrupole
coupling between the disk's mass quadrupole moments. These terms are of
course small but they might become important when extending the
analysis to include first order post-Newtonian corrections. By investigating
post-Newtonian corrections we hope to get a
better intuition for a semi-analytical treatment of the much more
complicated analysis of inspiraling NS-NS binaries.

\subsection*{Acknowledgements}
I thank G. Sch\"afer for helpful discussions and
encouragement. Conversations with A. Gopakumar are thankfully
acknowledged. This work was supported by the Deutsche
Forschungsgemeinschaft (DFG) through SFB/TR7 Gravitationswellenastronomie.

\appendix
\section{The quadrupole tensor of the disk and it's derivatives}

To calculate the quadrupole tensor of the disk, we start with the
general definition
\begin{eqnarray*}
    Q_{ij}=\int d^3x \rho(x^ix^j-\frac{1}{3}\delta^{ij}\vec r^2).
\end{eqnarray*}
Since $\rho_d$, the density of the disk, is confined to the $z=0$
plane and depends on $r$, the radial coordinate, only, the nondiagonal
elements of $Q_{ij}^{(d)}$ vanish. It is now easy to see that
\begin{align}
\label{Qdisk1}
     Q_{11}^{(d)}=\int r dr\, d\vp\, \rho(r)
     \frac{r^2}{3}(2\cos^2\vp-\sin^2\vp)=\frac{M_d}{15}\rd^2 =\frac{\al}{6}\rd^2.
\end{align}
Due to the symmetry it is clear that the relations 
\begin{align}
\label{Qdisk2}
      Q_{11}^{(d)}=Q_{22}^{(d)},\qquad Q_{33}^{(d)}=-2Q_{11}^{(d)}
\end{align}
hold. The second and third time derivative of $Q_{11}^{(d)}$ can be
calculated using the equation of motion Eq. (\ref{disk8}). With
\begin{align*}
     \dot{r}_d^2&=\frac{2E_d}{\al}+\frac{4C}{\rd}-\frac{h^2}{\rd^2}\\
     \ddot{r}_d&=\frac{h^2}{\rd^3}-\frac{2C}{\rd^2}=\frac{p_\vp^2}{\al^2\rd^3}-\frac{2C}{\rd^2}
\end{align*}
we obtain
\begin{align}
\label{Qdisk3}
      \ddot{Q}_{11}^{(d)}&=\frac{\al}{3}(\dot{r}_d^2+\rd\ddot{r}_d)
      =\frac{2\al}{3}\left(\frac{E_d}{\al}+\frac{C}{\rd}\right) \\
\label{Qdisk4}
      &=\frac{\al}{3}\left(\frac{p_r^2}{\al^2}+\frac{p_\vp^2}{\al^2\rd^2} -\frac{2C}{\rd}\right)=\frac{1}{3}\left[\frac{1}{\al}\left(p_r^2+\frac{p_\vp^2}{\rd^2}\right)-\frac{2C\al}{\rd}\right]
\end{align}
and
\begin{align}
\label{Qdisk5}
     \dddot{Q}^{(d)}_{11}=-\frac{2}{3}\frac{Cp_r}{\rd^2}.
\end{align}

\section{Time derivatives of the orbital quadrupole tensor}

By definition, the $z=0$ plane is the orbital plane. The $Q_{13}^{orb}$
and $Q_{23}^{orb}$ components of the orbital quadrupole tensor thus
vanish and we are left with
\begin{align}
     Q_{11}^{(o)}&=\frac{\mu R^2}{6}[1+3\cos(2\Phi)],\\
     Q_{22}^{(o)}&=\frac{\mu R^2}{6}[1-3\cos(2\Phi)],\\
     Q_{33}^{(o)}&=-\frac{\mu R^2}{3},\qquad
     Q_{12}^{(o)}=\frac{\mu R^2}{2}\sin(2\Phi).
\end{align}
The time derivatives are calculated using the (Newtonian) equations of
motion,
\begin{eqnarray*}
    \dot{R}^2=\frac{2E_{orb}}{\mu}+\frac{2G\M}{R}-R^2\dot{\Phi}^2,\qquad
    \ddot{R}=R\dot{\Phi}^2-\frac{G\M}{R^2},\qquad
    \ddot{\Phi}=-\frac{2}{R}\dot{R}\dot{\Phi}.
\end{eqnarray*}
One finds
\begin{align*}
      \ddot{Q}^{(o)}_{11}&=\frac{1}{3}\left[2E^{(o)}+\frac{G\M\mu}{R}\right]
      +\cos(2\Phi)\left[2E^{(o)}+\frac{G\M\mu}{R}-2\mu R^2\dot{\Phi}^2\right]
      -2\mu R\dot{R}\dot{\Phi}\sin(2\Phi) \nonumber \\
      &=\frac{1}{3}\left[\frac{P_R^2}{\mu}+\frac{P_\Phi^2}{\mu
      R^2}-\frac{G\M\mu}{R}\right] +
      \cos(2\Phi)\left[\frac{P_R^2}{\mu}-\frac{P_\Phi^2}{\mu
      R^2}-\frac{G\M\mu}{R}\right]-\frac{2P_RP_\Phi}{\mu
      R}\sin(2\Phi),\\
      \ddot{Q}^{(o)}_{22}&=\frac{1}{3}\left[2E^{(o)}+\frac{G\M\mu}{R}\right]
      -\cos(2\Phi)\left[2E^{(o)}+\frac{G\M\mu}{R}-2\mu R^2\dot{\Phi}^2\right]
      +2\mu R\dot{R}\dot{\Phi}\sin(2\Phi) \nonumber \\
      &=\frac{1}{3}\left[\frac{P_R^2}{\mu}+\frac{P_\Phi^2}{\mu
      R^2}-\frac{G\M\mu}{R}\right] -
      \cos(2\Phi)\left[\frac{P_R^2}{\mu}-\frac{P_\Phi^2}{\mu
      R^2}-\frac{G\M\mu}{R}\right]+\frac{2P_RP_\Phi}{\mu
      R}\sin(2\Phi),\\
      \ddot{Q}^{(o)}_{12}&=\left[2E^{(o)}+\frac{G\M\mu}{R}-2\mu
      R^2\dot{\Phi}^2\right]\sin(2\Phi)+2\mu
      R\dot{R}\dot{\Phi}\cos(2\Phi)\\
      &=\left[\frac{P_R^2}{\mu}-\frac{P_\Phi^2}{\mu
      R^2}-\frac{G\M\mu}{R}\right]\sin(2\Phi) +\frac{2P_RP_\Phi}{\mu
      R}\cos(2\Phi) \\
      \ddot{Q}_{33}^{(o)}&=-\frac{2}{3}\left[2E^{(o)}+\frac{G\M
      \mu}{R}\right] \\
      &=-\frac{2}{3}\left[\frac{P_R^2}{\mu}+\frac{P_\Phi^2}{\mu R^2}-\frac{G\M\mu}{R}\right].
\end{align*}
and
\begin{align*}
    \dddot{Q}_{11}^{(o)}&=-\frac{G\M
    P_R}{3R^2}\left[1+3\cos(2\Phi)\right]+\frac{4G\M
    P_\Phi}{R^3}\sin(2\Phi),\\
    \dddot{Q}^{(o)}_{22}&=-\frac{G\M
    P_R}{3R^2}\left[1-3\cos(2\Phi)\right] -\frac{4G\M
    P_\Phi}{R^3}\sin(2\Phi),\\
    \dddot{Q}_{12}^{(o)}&=-\frac{4G\M
    P_\Phi}{R^3}\cos(2\Phi)-\frac{G\M P_R}{R^2}\sin(2\Phi),\\
    \dddot{Q}^{(o)}_{33}&=\frac{2}{3}\frac{G\M P_R}{R^2}.
\end{align*}


\begin{figure}[h]
\begin{minipage}{0.46\linewidth}
\psfrag{y}{y}
\psfrag{in}{}
\psfrag{GMc2}{\hspace{-0.3cm}[$G\M/c^2$]}
\psfrag{x}{x}
\includegraphics[width=\linewidth]{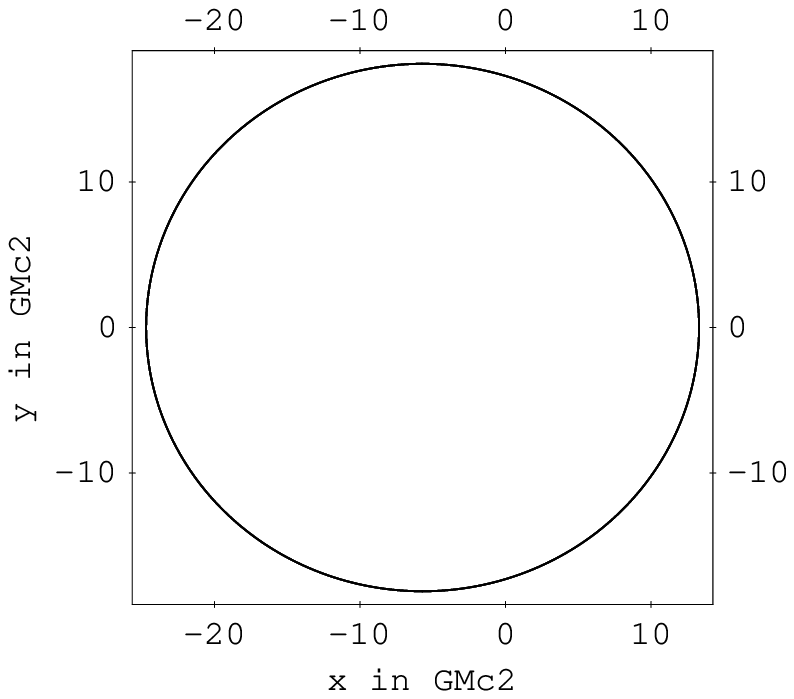}
\end{minipage}
\hfill
\begin{minipage}{0.46\linewidth}
\psfrag{y}{y}
\psfrag{in}{}
\psfrag{GMc2}{\hspace{-0.3cm}$[G\M/c^2$]}
\psfrag{x}{x}
\includegraphics[width=\linewidth]{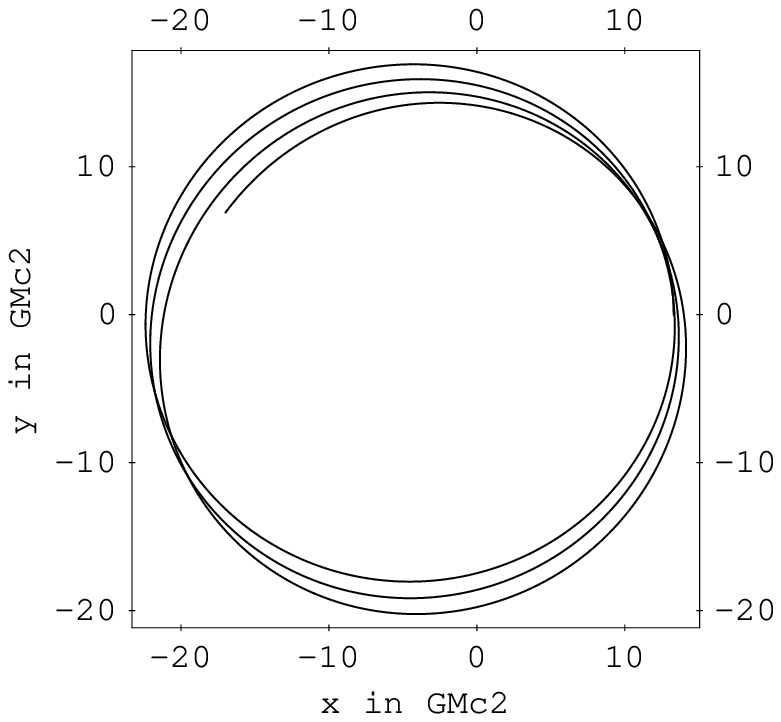}
\end{minipage}
\begin{minipage}{0.46\linewidth}
\psfrag{y}{y}
\psfrag{in}{}
\psfrag{GMc2}{\hspace{-0.3cm}[$G\M/c^2$]}
\psfrag{x}{x}
\includegraphics[width=\linewidth]{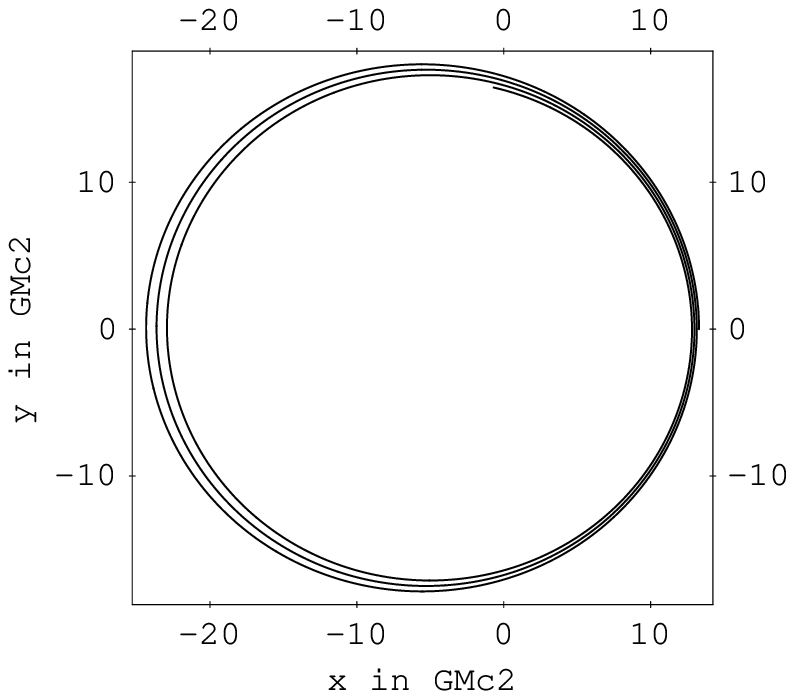}
\end{minipage}
\hfill
\begin{minipage}{0.46\linewidth}
\psfrag{y}{y}
\psfrag{in}{}
\psfrag{GMc2}{\hspace{-0.3cm}[$G\M/c^2$]}
\psfrag{x}{x}
\includegraphics[width=\linewidth]{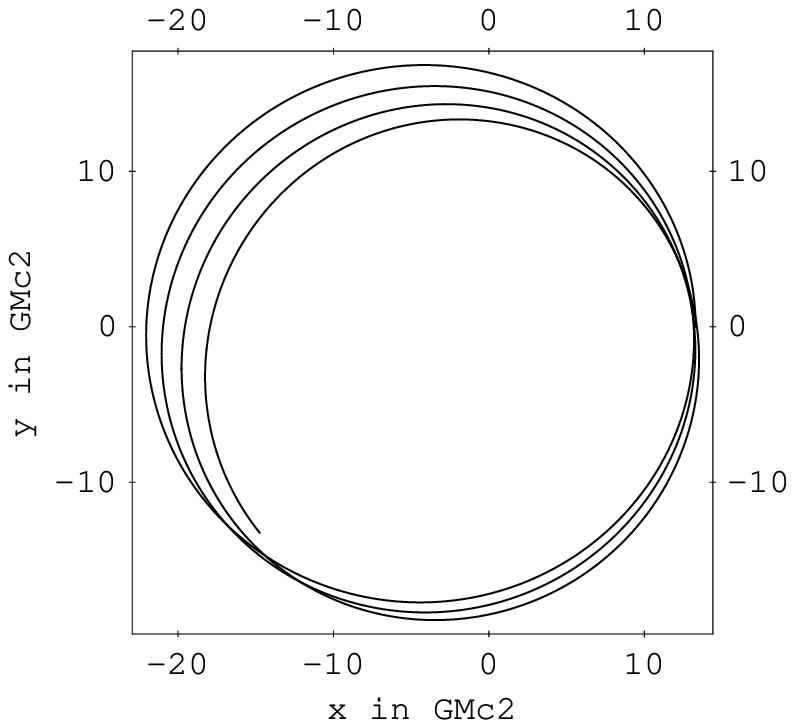}
\end{minipage}
\caption{Tidal interaction and gravitational radiation effects are
shown for a slightly elliptic $(e(0)=3/10$) binary with $a'(0)=19,
x_d(0)=X_d=6, M_d=M_p=1/2, \epsilon=1/10$. At $t=0$ the disk to
orbital frequency relation is $\omega_d/\omega_o =7.05$. The Newtonian
point-particle orbit is given  in the upper left figure.  Including tidal
interaction leads to a  periastron shift (upper right).
The figures on the bottom are calculated including leading order
gravitational radiation effects in the equations of motion.  Tidal
interaction speeds up the inspiralling process as can be seen by
comparing the orbit including (right) and excluding (left) tidal interaction.}
\label{fig:2}
\end{figure}

\begin{figure}[h]
\begin{minipage}{0.55\linewidth}
\psfrag{tT0}{$t/T_0$}
\psfrag{I22}{\hspace{0.5cm}$\ddot{I}^{22}$}
\psfrag{in}{}
\psfrag{muc2}{[$\mu c^2$]}
\includegraphics[width=\linewidth]{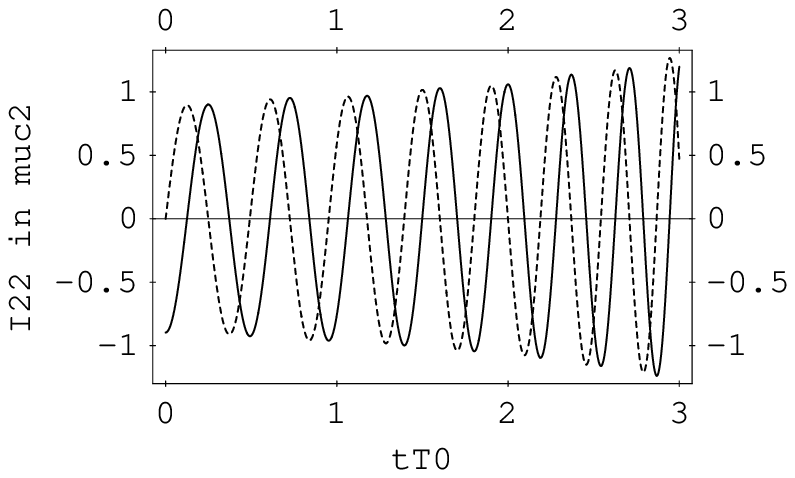}
\end{minipage}
\hfill
\begin{minipage}{0.55\linewidth}
\psfrag{tT0}{$t/T_0$}
\psfrag{I22}{\hspace{0.5cm}$\ddot{I}^{22}$}
\psfrag{in}{}
\psfrag{muc2}{$[\mu c^2]$}
\includegraphics[width=\linewidth]{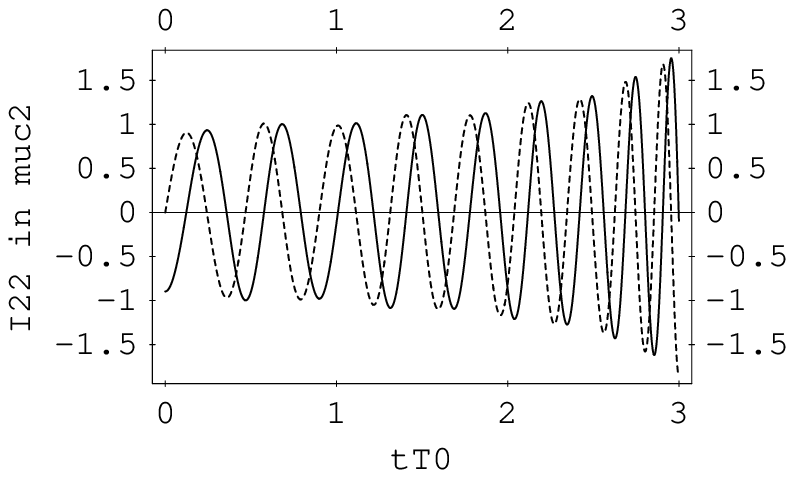}
\end{minipage}
\caption{The influence of the tidal interaction on the $\ddot{I}^{22}_{total}$
component of the binary's gravitational radiation field is shown.
Real and imaginary parts of $\ddot{I}^{22}_{total}$ are given without
(left) and including (right) tidal interaction.
In order to make the influence of tidal interaction on the
gravitational waveform  more
visible, the orbital separation is taken to be small ($M_d=M_p=1/2,
e(0)=0, a'(0)=10, x_d(0)=X_d=3, \epsilon=1/10$). The time is measured
in units of $T_0$, the initial orbital period.}
\label{fig:3}
\end{figure}

\begin{figure}[h]
\begin{minipage}{0.55\linewidth}
\psfrag{I20}{\hspace{0.5cm}$\ddot{I}^{20}$}
\psfrag{tT0}{$t/T_0$}
\psfrag{in}{}
\psfrag{muc2}{$[\mu c^2$]}
\includegraphics[width=\linewidth]{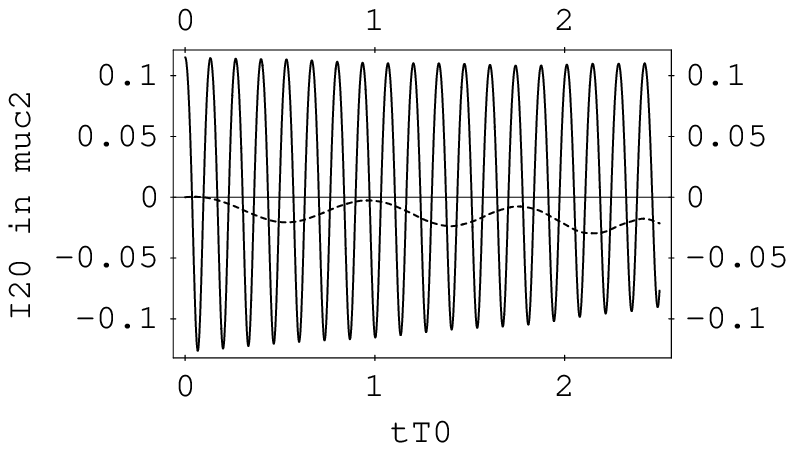}
\end{minipage}
\hfill
\begin{minipage}{0.55\linewidth}
\psfrag{I20total}{\hspace{0.9cm}$\ddot{I}^{20}_{total}$ }
\psfrag{tT0}{$t/T_0$}
\psfrag{in}{}
\psfrag{muc2}{\hspace{-0.4cm}$[\mu c^2]$}
\includegraphics[width=\linewidth]{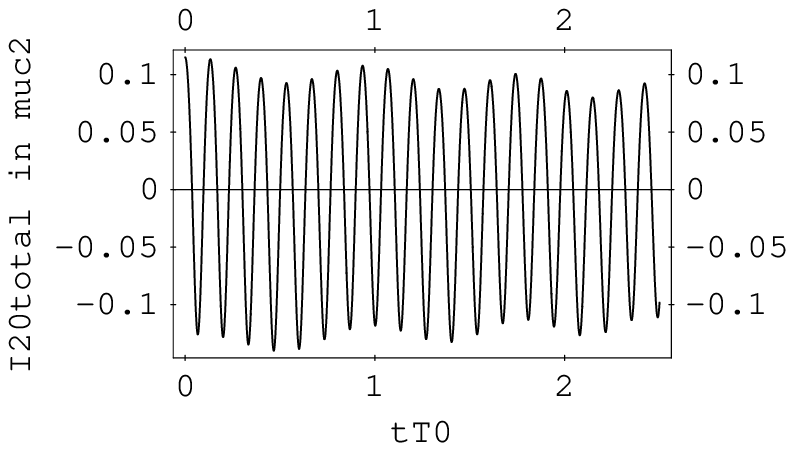}
\end{minipage}
\caption{$\ddot{I}^{20}_d$ (solid line) and $\ddot{I}^{20}_{orb}$
(dotted line) for the tidally coupled
system (left), and 
the $\ddot{I}^{20}_{total}$ component of the binary's leading
order gravitational waveform (right). The parameters are
the same as in Fig. 3. Although
the unperturbed orbit is assumed to be circular, 
$\ddot{I}^{20}_{orb}$ is \emph{not} zero due to tidal
interaction. $\ddot{I}^{20}_{total}$ is dominated by
the disk's contribution. The time is measured in units of $T_0$, the
initial orbital period.}
\label{fig:4}
\end{figure}


\begin{figure}
\begin{minipage}{0.55\linewidth}
\psfrag{I20}{\hspace{0.5cm}$\ddot{I}^{20}_d$}
\psfrag{in}{}
\psfrag{muc2}{[$\mu c^2$]}
\psfrag{tT0}{$t/T_0$}
\includegraphics[width=\linewidth]{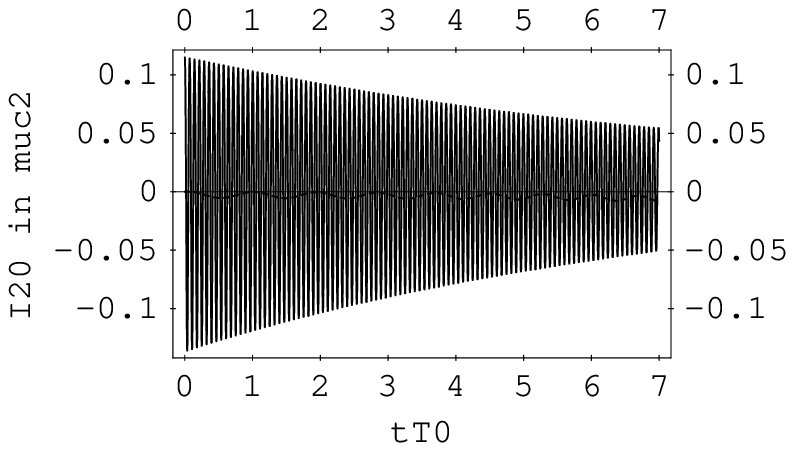}
\end{minipage}
\hfill
\begin{minipage}{0.55\linewidth}
\psfrag{I20tot}{\hspace{0.5cm}$\ddot{I}^{20}_{total}$}
\psfrag{in}{}
\psfrag{muc2}{\hspace{-0.4cm}[$\mu c^2$]}
\psfrag{tT0}{$t/T_0$}
\includegraphics[width=\linewidth]{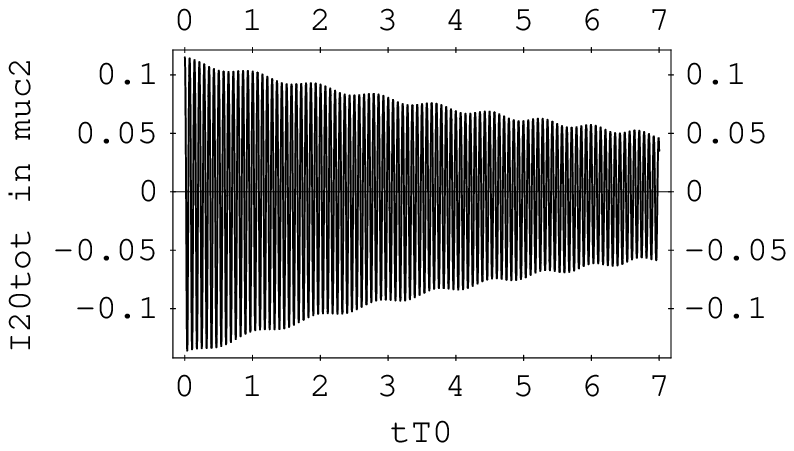}
\end{minipage}
\caption{The $\ddot{I}^{20}$ component of the leading order
gravitational radiation field of the binary for $a'(0)=15$, all other
parameters are the same as in figures~(\ref{fig:3}), (\ref{fig:4}). On
the left, the contributions of the disk is given, the right figure shows
$\ddot{I}^{20}_{total}$, which is dominated by the disk's
contribution. The time is measured in units of the inital orbital
period $T_0$.}
\label{fig:6}
\end{figure}

\begin{figure}[h]
\begin{center}
\psfrag{tT0}{$t/T_0$}
\psfrag{I22}{\hspace{0.6cm}$\ddot{I}^{22}$}
\psfrag{in}{}
\psfrag{muc2}{\hspace{-0.3cm}[$\mu c^2$]}
\includegraphics[width=10cm]{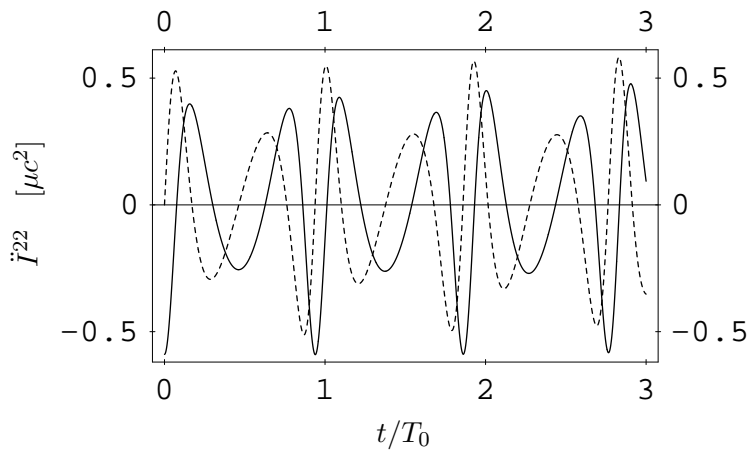}
\caption{Real (solid line)  and imaginary (dotted line) part of
$\ddot{I}^{22}_{total}$ for an elliptic
orbit with $e(0)=3/10$, $a'(0)=25, x_d(0)=X_d=6, \epsilon=1/10,
M_d=M_p=1/2$. Time is measured in fraction of the initial orbital
period $T_0$.}
\label{fig:7}
\end{center}
\end{figure}

\begin{figure}
\begin{center}
\begin{minipage}{0.7\linewidth}
\psfrag{tT0}{\large{$t/T_0$}}
\psfrag{I20}{\hspace{1cm}\large{$\ddot{I}^{20}$}}
\psfrag{in}{}
\psfrag{muc2}{\large{[$\mu c^2$]}}
\includegraphics[width=\linewidth]{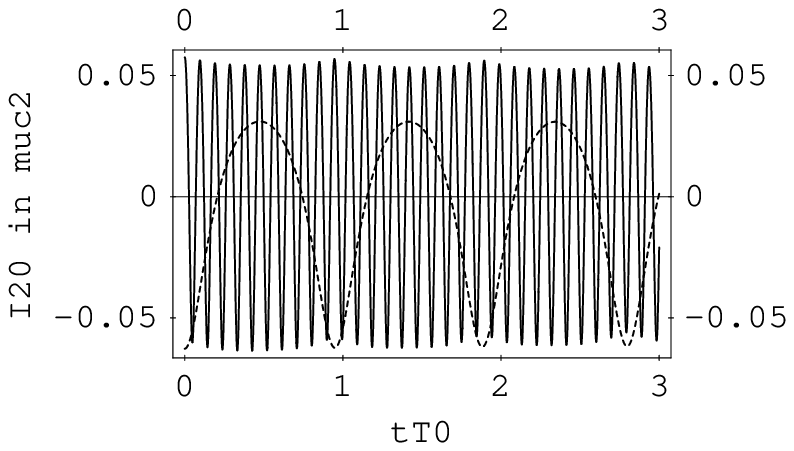}
\end{minipage}
\hfill
\begin{minipage}{0.7\linewidth}
\psfrag{I20tot}{\hspace{1cm}\large{$\ddot{I}^{20}_{total}$}}
\psfrag{in}{}
\psfrag{muc2}{\hspace{-0.4cm}\large{[$\mu c^2$]}}
\psfrag{tT0}{\large{$t/T_0$}}
\includegraphics[width=\linewidth]{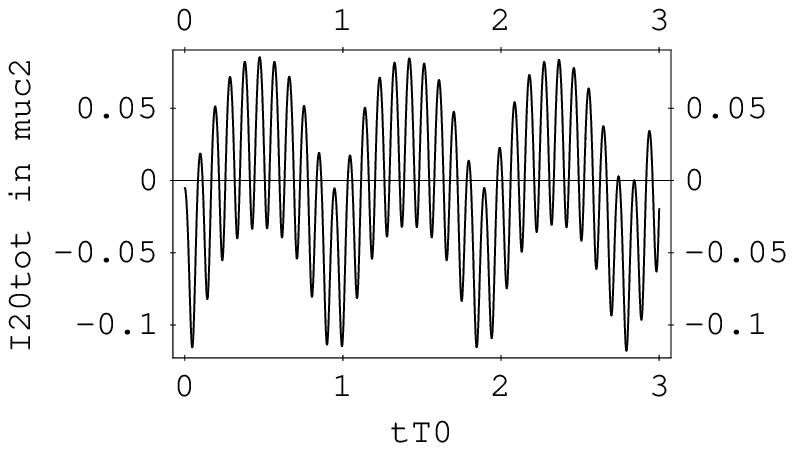}
\end{minipage}
\caption{$\ddot{I}^{20}$ for an elliptic orbit ($e(0)=0.3$).
In the left graph orbital (dotted line) and disk (solid line) contributions 
are given separately, the right graph shows the $\ddot{I}^{20}_{total}$
part of the binary's  leading  order gravitational waveform. The time
is given in fraction of the initial orbital period $T_0$.}
\label{fig:8}
\end{center}
\end{figure}

\begin{figure}
\begin{center}
\begin{minipage}{0.8\linewidth}
\psfrag{in}{}
\psfrag{tT0}{\large{$t/T_0$}}
\psfrag{I20di}{\hspace{1cm}\large{$\ddot{I}^{20}_{disk}$}}
\psfrag{muc2}{\hspace{-0.5cm}\large{[$\mu c^2$]}}
\includegraphics[width=\linewidth]{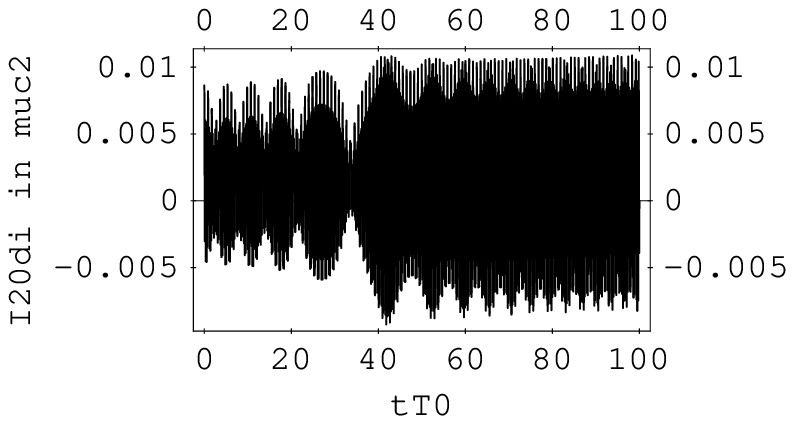}
\end{minipage}
\hfill
\begin{minipage}{0.8\linewidth}
\psfrag{I20or}{\hspace{1cm}\large{$\ddot{I}^{20}_{orb}$}}
\psfrag{muc2}{\hspace{-0.5cm}\large{[$\mu c^2$]}}
\psfrag{in}{}
\psfrag{tT0}{\large{$t/T_0$}}
\includegraphics[width=\linewidth]{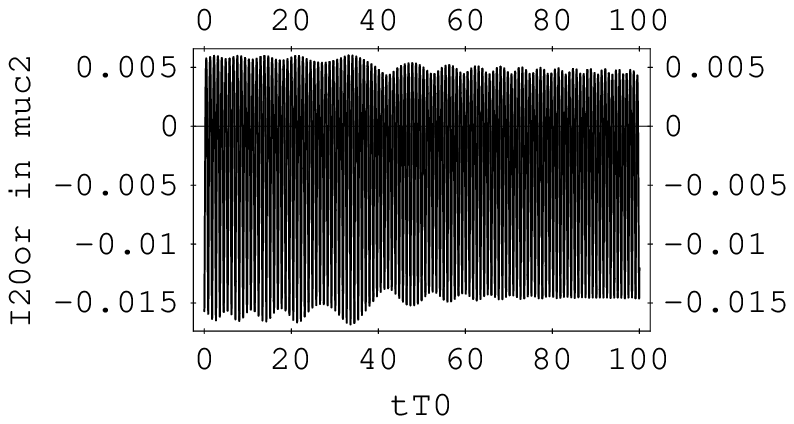}
\end{minipage}
\begin{minipage}{0.8\linewidth}
\psfrag{I20to}{\hspace{1cm}\large{$\ddot{I}^{20}_{total}$}}
\psfrag{muc2}{\hspace{-0.5cm}\large{[$\mu c^2$]}}
\psfrag{tT0}{\large{$t/T_0$}}
\psfrag{in}{}
\includegraphics[width=\linewidth]{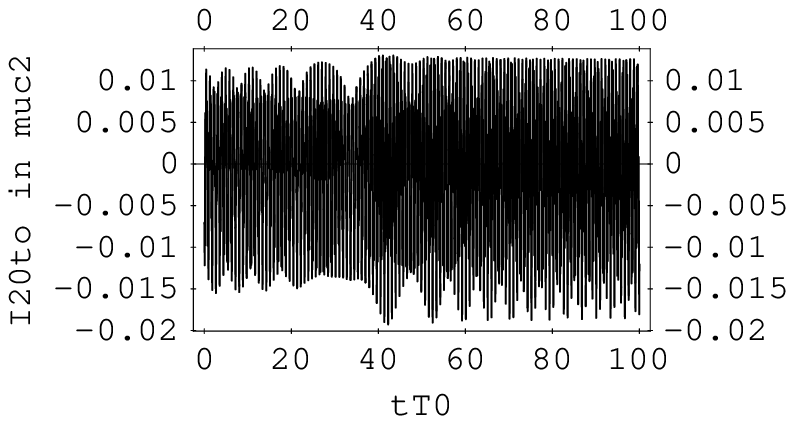}
\end{minipage}
\caption{The $\ddot{I}^{20}$ contribution to the leading order
gravitational wave radiation is shown for a binary with $a(0)=100,
e(0)=0.3, \eps(0)=0.1, x_d(0)= X_d =40, M_p=M_d=1/2$. The system
starts at $\omega_d/\omega_0=4.95$. The upper left figure shows
$\ddot{I}^{20}_{disk}$, while $\ddot{I}^{20}_{orb}$ is given in the
upper right figure. The total $\ddot{I}^{20}$ contribution is shown in
the last figure. The time is measured in units of the initial orbital
period $T_0$.}
\label{fig:9}
\end{center}
\end{figure}

\end{document}